\documentclass[
11pt, 
english, 
onehalfspacing, 
headsepline, 
]{MastersDoctoralThesis} 

\usepackage[utf8]{inputenc} 
\usepackage[T1]{fontenc} 

\usepackage{mathpazo} 

\usepackage[backend=bibtex,style=authoryear,natbib=true]{biblatex} 

\usepackage[skip=10pt plus1pt, indent=15pt]{parskip}
\usepackage{indentfirst}
\usepackage{pgf-pie}
\usepackage{pgfplots}
\usepackage{epigraph}

\usepackage[autostyle=true]{csquotes} 

\thesistitle{Updated Standard for Secure Satellite Communications: Analysis of Satellites, Attack Vectors, Existing Standards and, Enterprise and Security Architectures} 
\supervisor{Dr. Santanu \textsc{Dash}} 
\examiner{} 
\degree{MSc in Information Security} 

\author{Rupok Chowdhury Protik} 
\addresses{} 

\subject{Information Security} 
\university{\href{https://www.royalholloway.ac.uk/}{Royal Holloway, University of London}} 
\department{\href{https://www.royalholloway.ac.uk/research-and-teaching/departments-and-schools/information-security/}{Department of Information Security}} 
\group{\href{https://intranet.royalholloway.ac.uk/students/information-hub/academic/school-of-engineering-physical-and-mathematical-sciences.aspx}{School of Engineering, Physical and Mathematical Sciences}} 
\faculty{Dr. Santanu Dash} 

\AtBeginDocument{
\hypersetup{pdftitle=\ttitle} 
\hypersetup{pdfauthor=\authorname} 
\hypersetup{pdfkeywords=\keywordnames} 
\hypersetup{colorlinks, allcolors=orange, urlcolor=orange,} 
}

\begin{document}

\frontmatter 

\pagestyle{plain} 


\begin{titlepage}
\begin{center}

\vspace*{.06\textheight}
{\scshape\LARGE \univname\par}\vspace{1.5cm} 
\textsc{\Large MSc. Project}\\[0.5cm] 

\HRule \\[0.4cm] 
{\huge \bfseries \ttitle\par}\vspace{0.4cm} 
\HRule \\[1.5cm] 
 
\begin{minipage}[t]{1\textwidth}
\centering
\emph{Author:}\\
{\authorname} 
\end{minipage}
\begin{minipage}[t]{0.4\textwidth}
\begin{flushright} \large
\end{flushright}
\end{minipage}\\[3cm]
 
\vfill

\large \textit{Submitted as part of the requirements for the award of the \\
\degreename}\\[0.3cm] 
\textit{at}\\[0.4cm]
\univname\\[2cm] 
\vfill

{\large \today}\\[4cm] 

\vfill
\end{center}
\end{titlepage}


\begin{abstract}
\addchaptertocentry{Executive Summary} 
Satellites play a vital role in remote communication where traditional communication mediums struggle to provide benefits over associated costs and efficiency. In recent years, satellite communication has achieved utter interest in the industry due to the achievement of high data rates through the massive deployment of LEO satellites. Because of the complex diversity in types of satellites, communication methodologies, technological obstacles, environmental limitations, elements in the entire ecosystem, massive financial impact, geopolitical conflict and domination, easier access to satellite communications, and various other reasons, the threat vectors are rising in the threat landscape. To achieve resilience against those, only technological solutions are not enough. An effective approach will be through security standards. However, there is a considerable gap in the industry regarding a generic security standard framework for satellite communication and space data systems. A few countries and space agencies have their own standard framework and private policies. However, many of those are either private, serve the specific requirements of specific missions, or have not been updated for a long time.

This project report will focus on identifying, categorizing, comparing, and assessing elements, threat landscape, enterprise security architectures, and available public standards of satellite communication and space data systems. After that, it will utilize the knowledge to propose an updated standard framework for secure satellite communications and space data systems.
\end{abstract}


\tableofcontents 

\listoffigures 

\listoftables 


\begin{abbreviations}{ll} 

\textbf{ADM} & \textbf{A}rchitecture \textbf{D}evelopment \textbf{M}ethod\\
\textbf{ALMV} & \textbf{A}ir \textbf{L}aunched \textbf{M}iniature \textbf{V}ehicle\\
\textbf{ASAT} & \textbf{A}nti \textbf{Sat}ellite\\
\textbf{BISS} & \textbf{B}asic \textbf{I}nteroperable \textbf{S}crambling \textbf{S}ystem\\
\textbf{BSI} & \textbf{B}ritish \textbf{S}tandards \textbf{I}nstitute\\
\textbf{BSS} & \textbf{B}roadcasting \textbf{S}atellite \textbf{S}ervice\\
\textbf{CAGR} & \textbf{C}ompound \textbf{A}nnual \textbf{G}rowth \textbf{R}ate\\
\textbf{CCSDS} & \textbf{C}onsultative \textbf{C}ommittee for \textbf{Sp}ace \textbf{D}ata \textbf{S}ystems\\
\textbf{CNSSI} & \textbf{C}ommittee on \textbf{N}ational \textbf{S}ecurity \textbf{S}ystems \textbf{I}nstruction\\
\textbf{FSS} & \textbf{F}ixed \textbf{S}atellite \textbf{S}ervice\\
\textbf{GEO} & \textbf{G}eosynchronous \textbf{E}quatorial \textbf{O}rbit\\
\textbf{GNSS} & \textbf{G}lobal \textbf{N}avigation \textbf{S}atellite \textbf{S}ystems\\
\textbf{GPS} & \textbf{G}lobal \textbf{P}ositioning \textbf{S}ystem\\
\textbf{GTO} & \textbf{G}eostationary \textbf{T}ransfer \textbf{O}rbit\\
\textbf{HF} & \textbf{H}igh \textbf{F}requency\\
\textbf{HPM} & \textbf{H}igh \textbf{P}ower \textbf{M}icrowave\\
\textbf{IFFT} & \textbf{I}nverse \textbf{F}ast \textbf{F}ourier \textbf{T}ransform\\
\textbf{JTIDS} & \textbf{J}oint \textbf{T}actical \textbf{I}nformation \textbf{D}istribution \textbf{S}ystems\\
\textbf{LEO} & \textbf{L}ow \textbf{E}arth \textbf{O}rbit\\
\textbf{LOA} & \textbf{L}ift \textbf{O}ff \textbf{A}ligned\\
\textbf{LOD} & \textbf{L}ift \textbf{O}ff \textbf{D}elay\\
\textbf{MEO} & \textbf{M}edium \textbf{E}arth \textbf{O}rbit\\
\textbf{MIDS} & \textbf{M}ultifunctional \textbf{I}nformation \textbf{D}istribution \textbf{S}ystems\\
\textbf{MSS} & \textbf{M}obile \textbf{S}atellite \textbf{S}ervice\\
\textbf{NASA} & \textbf{N}ational \textbf{A}eronautics and \textbf{S}pace \textbf{A}dministration\\
\textbf{NCSC} & \textbf{N}ational \textbf{C}yber \textbf{S}ecurity \textbf{C}entre\\
\textbf{NIST} & \textbf{N}ational \textbf{I}nstitute of \textbf{S}tandards and \textbf{T}echnology\\
\textbf{PDT} & \textbf{P}ayload \textbf{D}ata \textbf{T}ransmission\\
\textbf{PPD} & \textbf{P}ersonal \textbf{P}rivacy \textbf{D}evices\\
\textbf{PRN} & \textbf{P}seudo \textbf{R}andom \textbf{N}oise\\
\textbf{SCORE} & \textbf{S}ignal \textbf{C}ommunication by \textbf{O}rbiting \textbf{R}elay \textbf{E}quipment\\
\textbf{SDR} & \textbf{S}oftware \textbf{D}efined \textbf{R}adio\\
\textbf{SSO} & \textbf{S}un-\textbf{S}ynchronous \textbf{O}rbit\\
\textbf{STFT} & \textbf{S}hort \textbf{T}ime \textbf{F}ourier \textbf{T}ransform\\
\textbf{TT\&C} & \textbf{T}elemetry \textbf{T}racking and \textbf{C}ommand\\
\textbf{UHF} & \textbf{U}ltra \textbf{H}igh \textbf{F}requency\\
\textbf{VHF} & \textbf{V}ery \textbf{H}igh \textbf{F}requency\\
\textbf{VPN} & \textbf{V}irtual \textbf{P}rivate \textbf{N}etwork\\

\end{abbreviations}


\mainmatter 

\pagestyle{thesis} 



\chapter{Introduction} 

\label{Chapter1} 


\newcommand{\keyword}[1]{\textbf{#1}}
\newcommand{\tabhead}[1]{\textbf{#1}}
\newcommand{\code}[1]{\texttt{#1}}
\newcommand{\file}[1]{\texttt{\bfseries#1}}
\newcommand{\option}[1]{\texttt{\itshape#1}}


\section{Background}
Satellites are a crucial element in remote wireless communication where it is difficult to reach through other traditional communication mediums like cable or fiber. From a commercial vessel floating in the middle of the ocean to a personal internet connection, the demand for satellite communication is increasing day by day to gain better coverage and high-volume data transfer ability.

Satellites can be used for various purposes. The most common usage is for television broadcasting where a large coverage is necessary. With satellites, television programs can be broadcast over multiple countries and regions at the same time. Telecommunication is another significant usage of satellites. It makes telecommunication possible after a natural disaster when other communication mediums face failure. It also enables communication to areas that are extremely challenging to reach over conventional communication mediums like maritime ships or military vehicles in the middle of the ocean. If there is a clear path between the telecommunication user and the satellite (as the microwave energy travels in a straight line), a close connection can be established without requiring any repeaters to boost the signal in between. It removes the necessity of installing repeaters to maintain the line-of-sight for sending or receiving microwave signals to/from large distances.

A prevalent use of satellites is providing broadband internet connectivity. In long-distance cable connections, specific repeaters have to be deployed to amply the signal due to the energy loss during signal transfer. This is financially not profitable while establishing a connection to remote locations. Besides communicating over satellites, maritime vessels also heavily use satellites for GPS-based navigation.

Another significant usage of satellites is collecting weather information from space. It enables monitoring weather conditions from a global perspective. Satellites are also an important source of environmental information. Satellites orbiting the earth can continuously send environmental and climate information which can be stored and analyzed to predict climate changes in the future and identify patterns in climate change. Another significant arena of satellites is space research. Before launching satellites, all space observation was done from ground telescopes, radars, and other ground-based specialized hardware receivers. Up until today, more than 50 satellites have been launched into space which were used for collecting information about space. The two most significant satellite telescopes are the Hubble Space Telescope (\cite{noauthor_hubble_nodate}) and the James Webb Space Telescope (\cite{noauthor_james_nodate}) which have immense contributions to the space knowledge we have gathered up until now. Militaries of different countries also heavily use satellite-based communication for collecting real-time accurate intelligence from their field agents. They also use satellites for accurately guiding weapons and aircraft from a long distance.

\begin{figure}
    \centering
        \includegraphics[scale=0.7]{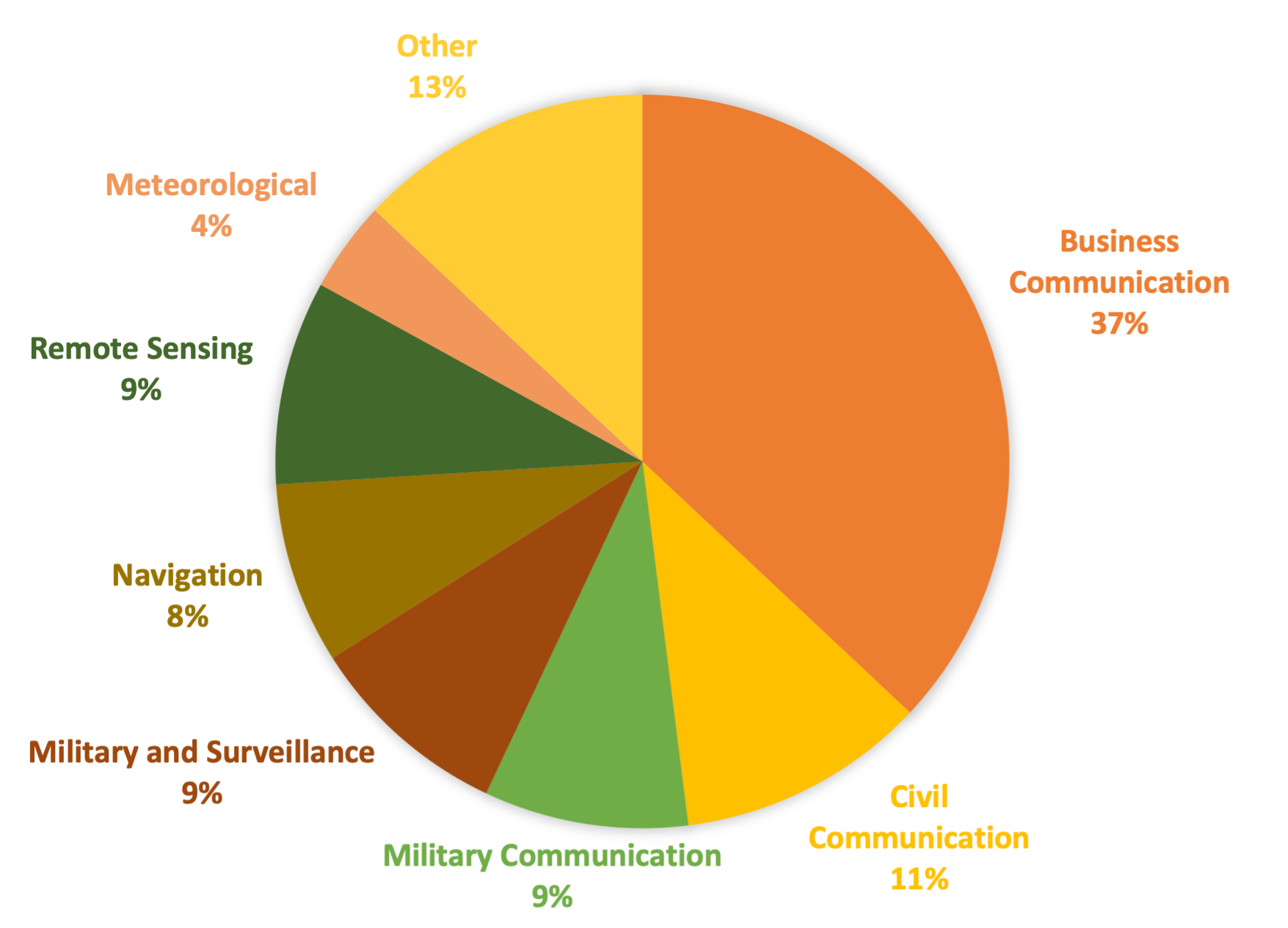}
        \caption{Share of satellite usage by various industries}
        \label{fig:share_of_usage_percentage}
\end{figure}

On top of everything, to utilize the advantages of satellites, organizations do not need to deploy their own satellite in space by investing a significantly large amount of money. Rather, smaller organizations, even individuals, can rent the satellite capacity from other organizations that have deployed satellites up in space. Ground stations or receivers can be deployed at multiple locations very quickly and cost-effectively to establish a connection to the satellites. The receivers today can be as small as a handheld device. A majority of the satellites in space are basically microwave signal repeaters that relay the signal sent from ground-based stations. However, observation and other types of satellites generate information and then send it to the ground stations instead of relays. Jeffrey Bardin (\cite{bardin_satellite_2013}) identified that the largest usage of satellites is by business communication users. The other prominent users are the military and civil communication users. This is demonstrated in \ref{fig:share_of_usage_percentage}.

Each sector depending on Satellite communication has its own huge share in business. For example, the shipping industry alone is responsible for 90\% of the trades in the world (\cite{papadimitriou_space_2019}) which heavily depends on GNSS (\cite{ben_farah_cyber_2022}). As per the GrandViewResearch report (\cite{noauthor_satellite_nodate}), the overall market size of Satellite Communication in 2022 is USD 77.10 Billion, and consolidated revenue in Satellite Communication is expected to be USD 159.50 Billion. The CAGR is 9.5\% from the year 2022 to 2030 based on the historical data from the year 2017 to 2020. FortuneBusinessInsights also forecasted the same 9.54\% CAGR for the period of the year 2022 to 2029 (\cite{noauthor_satellite_nodate-1}). Besides government and military, telecommunication and broadcasting media sectors have been occupying a significant portion of the user base in satellite communication. To achieve global connectivity objectives, private business organizations are investing more in the satellite communication sector. In the 1990s, narrow-band Satellites were introduced and that initialized the commercialization and privatization of Satellite usage (\cite{han_recent_2022}). In today's world, satellite-based communication is considered to be a significant part of 5G advancement (B5G) and 6G communication standards (\cite{noauthor_space_nodate}).

Depending on the distance of the satellite from the earth's ground level, some satellites can cover up to one-third of the entire earth's surface area at the same time. For example, a Geostationary Satellite can have coverage over 40 degrees of latitude on Earth (\cite{peterson_satellite_2003}). Due to this unchallenged advantage, researchers and domain experts have always been working hard to further develop this communication medium. Since the beginning of the journey, satellite technologies kept evolving to meet the demand in the industry. With recent developments in technologies and advancement and with heavy investment from private business sectors, satellite communication now allows users to achieve gigabit speed in personal communication and transferring data through satellites without managing long cables and expensive infrastructure. The upcoming O3b mPOWER constellation will be able to provide up to multiple gigabytes per second to a single user (\cite{noauthor_boeing_nodate}). In the summer of 2021, NASA showcased the dynamic power of laser-based communication for satellites which will be capable of transmitting more data than ever before (\cite{noauthor_laser_nodate}).

\begin{figure}[h!]
    \centering
        \includegraphics[scale=1]{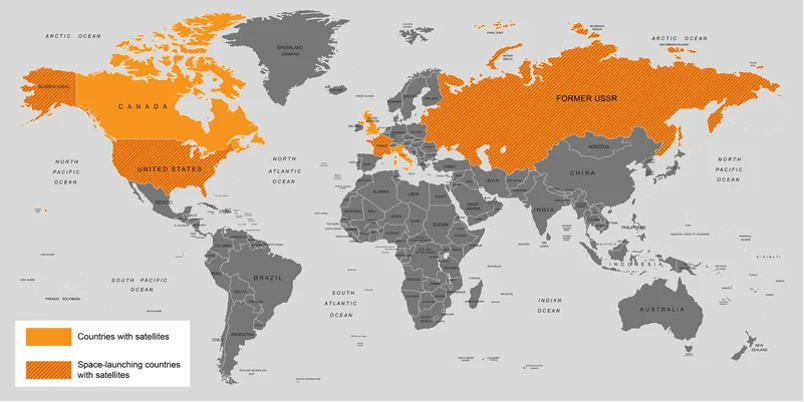}
        \caption{Countries involved with Satellites in year 1966 (\cite{noauthor_ucs_nodate})}
        \label{fig:satellite_countries_1966}
\end{figure}

\begin{figure}[h!]
    \centering
        \includegraphics[scale=1]{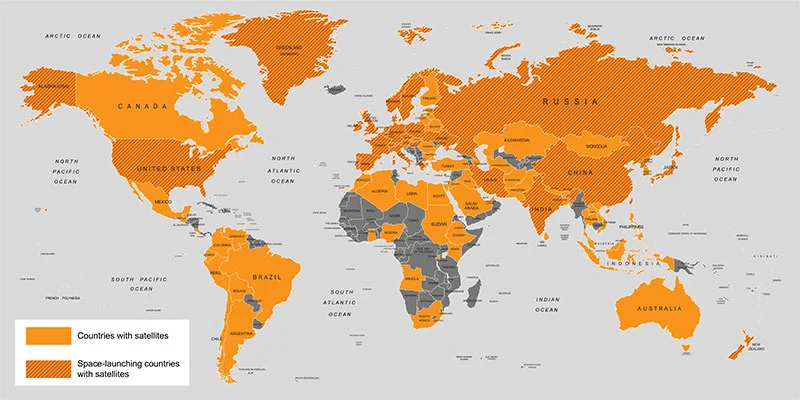}
        \caption{Countries involved with Satellites in year 2020 (\cite{noauthor_ucs_nodate})}
        \label{fig:satellite_countries_2020}
\end{figure}

\begin{figure}[ht]
    \centering
        \includegraphics[scale=0.34]{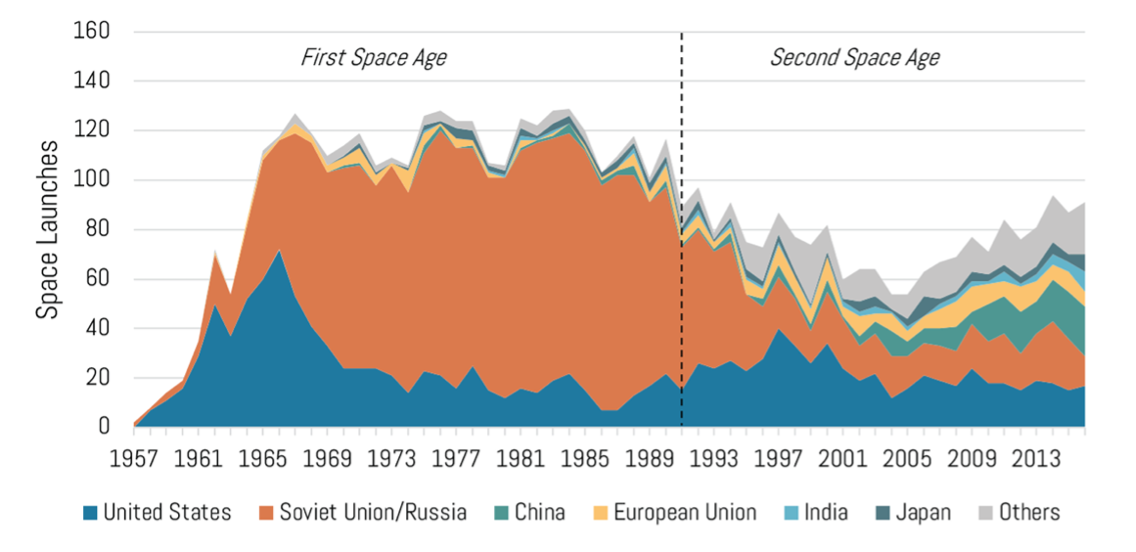}
        \caption{Satellites launched by countries until 2016 (\cite{spacetrackSpaceTrackorg})}
        \label{fig:satellite_launch_countries}
\end{figure}

\begin{figure}[htbp]
    \centering
        \includegraphics[scale=0.365]{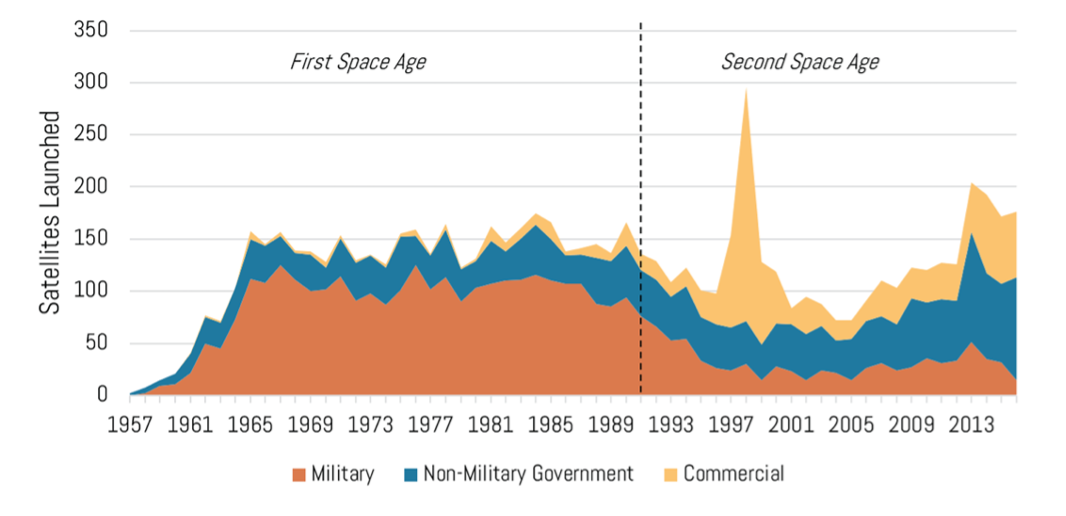}
        \caption{Satellites launched by types until 2016 (\cite{spacetrackSpaceTrackorg})}
        \label{fig:satellite_launch_type}
\end{figure}

As per the European Space Agency report released in 2022 (\cite{noauthor_esas_nodate}), a record number of rockets flew into space to put multiple satellites in orbit. This is likely to be increased in the upcoming year taking the market size expansion into consideration based on the CAGR. Since the first trial of Satellite Communication at the Tokyo Olympics in 1964 (\cite{han_recent_2022}), the use of Satellites kept increasing. More and more countries are becoming interested in Satellites. \ref{fig:satellite_countries_1966}, \ref{fig:satellite_countries_2020}, \ref{fig:satellite_launch_countries} and \ref{fig:satellite_launch_type} provides a clear representation of increased interest in satellite usage among state nations from the year 1966 to the year 2020.

\section{Motivation}

Based on the dependency on satellite communication and significant financial impact, an attack on satellite communication can have catastrophic adverse effects ranging from financial damage to the loss of individual lives. Because of the possible wideness of impacts and weakness in the legacy and costly infrastructure of satellite communications, adversaries are showing increased interest in attacking satellite communications. All three core security services – Confidentiality, Integrity, and Availability can be attacked in Satellite communications. If data is not encrypted with a strong algorithm, confidentiality can be impacted. If an appropriate cryptographic mechanism is not deployed, integrity can be affected. Recent research has proved that denial of service attack is possible, even on the latest technology implemented by the private commercial organization – SpaceX (\cite{smailes_dishing_2023}). Satellite communication utilizes the wireless communication link due to the architecture of this medium. Besides the common attacks on wireless networks, satellite communication is susceptible to spoofing, jamming signal-level attacks, data-level attacks (\cite{wu_spoofing_2020}), information-theoretic attacks, cryptographic attacks (\cite{tedeschi_satellite-based_2022}), physical layer attack, network layer attacks, denial of service attacks and many more. Several research over the past few decades have proposed different mitigation strategies for the aforementioned attacks. There is considerable scope for research for a generalized updated satellite security standard that can be incorporated by all parties involved in satellite communications. 

\paragraph{Government Interest and Author's Involvement} -- The author was privileged to be one of the 18 people from the whole United Kingdom invited by the NCSC UK for a workshop on satellite communication security. This resembles the interest of the government of the United Kingdom and the importance of securing satellite communications and portrays the impact of it in the upcoming decade. The author was highly encouraged to research further into this domain and utilized the knowledge achieved from the aforementioned workshop in this report.

\section{Objectives}
\begin{itemize}
  \item \textbf{Identification and Categorization of Satellites} -- This project will identify the categorization criteria, types, and elements of satellite communication from publicly available information. This will also identify the importance and impact of these on the security of satellite communications.
  \item \textbf{Communication Development Timeline} -- This report will research and create a timeline of the development of satellite communication technologies to get a better understanding of the growth of satellite communication technologies to develop an updated standard for achieving higher resilience in satellite communications.
  \item \textbf{Identification and Categorization of Threats} -- This report will research, analyze, and categorize the theoretical and practical threat vectors for satellite communications to define a precise scope for proposing a satellite communication security standard.
  \item \textbf{Existing Standard Analysis} -- This report will identify and analyze relevant public standards for secure satellite communications. The goal is to understand the current status of the standards and to find opportunities for advancement in this specific segment.
  \item \textbf{Propose Updated Standard} -- Finally this report will develop and propose an updated standard for secure satellite communications and space data systems based on the knowledge acquired through the previous segments of this report for achieving higher resilience against satellite communication threats.
\end{itemize}

\section{Structure of the Project}
At first, this project focused on a literature review of previous research and publications to develop an overall understanding of the proposed security solutions for securing satellite communications in Chapter \ref{Chapter1}. This report conveyed an open survey to consider the opinions of the domain experts and associates in the cyber security field in the same Chapter \ref{Chapter1}.

Understanding the communication over the satellites requires identifying and categorizing the types of satellites that are already deployed in space. This is discussed in Chapter \ref{Chapter2} of this report.

This report developed a timeline of satellite development from the beginning in the year 1945 until the year 2023. This is also presented in Chapter \ref{Chapter2}.

A grant chart of satellite deployment and their lifetime is demonstrated in Chapter \ref{Chapter2}.

Attacks on satellite communications are described in Chapter \ref{Chapter3}.

Analysis of existing significant standards is described in Chapter \ref{Chapter4}.

Before proceeding to developing and proposing an updated security framework for satellite communications and space data systems, this report analyzed the popular enterprise architectures and cyber security frameworks and models to find an effective framework or model to accommodate within the proposed framework in Chapter \ref{Chapter5}.

Then in Chapter \ref{Chapter6}, this report proposes an updated standard for secure satellite communications and space data systems and finally concludes the project in Chapter \ref{Chapter7}.

\section{Literature Review}
Satellite communication started way back in 1958 as part of a US Government project - SCORE (\cite{noauthor_satellite_nodate-4}). Since then, this communication medium has been used continuously through different other satellites and by satellite communication service providers including multiple governments and private organizations for various purposes including maritime communication, global positioning, weather forecasting, military communication, etc. (\cite{maral_satellite_2020}). The European Space Agency was considering the integration of satellite communication capabilities within the 5G spectrum (\cite{noauthor_esa_nodate}). Satellite communication will also play a vital role in the upcoming 6G communication (\cite{saeed_point--point_2021}). 5465 artificial satellites are orbiting around the earth as of April 30, 2022 (\cite{noauthor_number_nodate}). The European Space Agency categorized all these satellites into 6 major categories based on their orbits - GEO, LEO, MEO, SSO, GTO, and L-points (\cite{noauthor_types_nodate}). The satellite communication market is expected to reach USD 41.86 Billion by 2025 (\cite{tedeschi_satellite-based_2022}). Moreover, the value of the assets depending on GNSS alone is around 800 Billion Euros as per the European Commission survey (\cite{noauthor_global_2011}). Because of the financial impact, this communication medium has acquired increased attention from attackers.

According to James Turgal, the managing director of Cyber Risk Services at Deloitte, the human factor and the supply chain are the main areas of concern for cyber security around satellite communication (\cite{holmes_mark_growing_nodate}). That is why this report will analyze the possible and effective manners of securing satellite communication only from the ground station to the satellites but also will discuss the human factors revolving around the security in satellite communication. Different methods of cyber-attack are possible with the current satellite communication technology (\cite{santamarta_satcom_nodate}).

The current satellite communication has a wide variety of vulnerabilities including, but not limited to, backdoors, hardcoded credentials, insecure protocol, jamming, spoofing, and hijacking (\cite{rooker_satellite_2008}). Among them, some attacks are intentional, but some are unintentional. For example, some amateur radio services, terrestrial video broadcasting signals (DVB-T), multicarrier modulated satellite communication systems, etc. can cause out-of-band interference unintentionally. In-band unintentional interference can be caused by military JTIDS, MIDS, civilian radars (1215–1400 MHz), etc. Intentional interference can be caused by many devices including PPD used in car jammers (\cite{noauthor_global_2011}). No proven records of spoofing attacks by state nations have been reported as of 2011 but theoretically, it can be achieved with simple SDR and GNSS simulators. The core purpose of spoofing is to deceive the GNSS receiver's position and timing information. It can be achieved in many ways including, but not limited to, Navigation Message Attack, PRN Code Level Attack, LoD, LoA, Meaconing and selective delay, Jam, and Spoof, Noline of sight Spoofing, Trajectory spoofing, etc. (\cite{noauthor_global_2011}). Different mitigation plans have been proposed by different research for different kinds of attacks. Current mitigation techniques can be categorized into different categories including, but not limited to, time, frequency, time-frequency, and spatial-time domains (\cite{ioannides_known_2016}). To deal with the interference, the frequency domain mitigation approach removes the harmonic interference by eliminating the signal exceeding the detection threshold and then synthesizing the signal back to the original time-domain state (\cite{raimondi_mitigating_2006}). To mitigate the jamming attack, Wigner Ville distribution, STFT, and IFFT methods are utilized (\cite{zhang_anti-jamming_2001}).

However, previous research was on specific areas of the complex and large satellite communication systems instead of the security as a whole for the satellite communications and space data systems. This provides the opportunity to convey further research on a high-level security perspective for satellite communications.

\section{Research Methodology}
This project follows the "Analytical" approach described in the MSc. Information Security Project Handbook 2022/2023 version 1.2, Section 1.1. The project handbook mentions assessing standardization work under the "Analytical" paragraph in the aforementioned section.

\section{Survey Data Analysis}

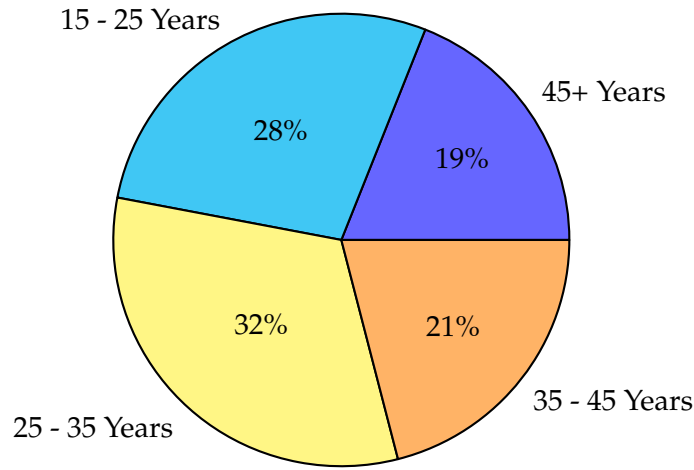
\begin{figure}
    \centering
      \begin{tikzpicture}
        \pie{19/45+ Years, 28/15 - 25 Years, 32/25 - 35 Years, 21/35 - 45 Years}
    \end{tikzpicture}  
    \caption{Survey Participants Age Groups}
    \label{fig:participants_age_group}
\end{figure}

This project developed a set of questionnaires to collect generic opinions regarding satellite communication security. The questions and many of the answer choices were intentionally kept open to make it more inclusive. This provided the opportunity to acquire feedback from people within a diversified demographic. The feedback was provided by people of:
\begin{itemize}
    \item Different age group
    \item Different educational qualifications
    \item Different organizations
    \item Different field of work domains
    \item Different countries
\end{itemize}

Among the participants, 67\% had at least one postgraduate degree, 26\% had an undergraduate degree, and 7\% either had a PhD degree or a current PhD student.

Age Group Analysis is presented in \ref{fig:participants_age_group}

\begin{itemize}
    \item 28\% were between 15 - 25 years old
    \item 32\% were between 25 - 35 years old
    \item 21\% were between 35 - 45 years old
    \item 19\% were above 45 years old
\end{itemize}

The participants were from 28 different organizations as specified in \ref{tab:participating_organizations}.

\begin{table}[h!]
\centering
\begin{tabular}{l l l}
\midrule
National Cyber Security Agency (NCSC) & Ministry of Defense (MoD), UK\\
GitHub (Microsoft) & KPMG\\
BP & Mission Aviation Fellowship (MAF)\\
BAE Systems & Liberty Global\\
Royal Holloway, University of London & Utrecht University\\
Northumbria University & University of Kent\\
University of Westminster & De Montfort University\\
Georgetown University & Coventry University\\
CDW, UK & Bylor LTD\\
Alstom & Gemeente Alphen aan den Rijn\\
Tonnik Consulting & Thought Machine\\
STM & Sky\\
Babylon Health, UK & Mambu\\
\bottomrule\\
\end{tabular}
\caption{Organizations of the participants in the survey}
\label{tab:participating_organizations}
\end{table}

Participants were from a diversified professional domain including – Cybersecurity, Information Security, Security and Compliance, Information Technology, Finance, University Lecturer, Mathematics, Business Information Systems, Business Management, Strategic Management, Engineering, Developer, and Software Architect.
The survey acquired feedback from participants within geographically diversified areas. It recorded responses from countries including England, Qatar, Germany, Netherlands, Northern Ireland, Greece, USA, India, and UAE.

To understand the participants' perception regarding dependence on satellite communication, the survey asked if they have traveled by air in the last 3 years. 86\% of the participants confirmed that they did and 14\% did not. This will help in understanding the following area where the participants were asked if they think they will be impacted if a major attack happens on satellite communication.

For a similar reason, participants were asked if they ordered any product that was shipped from abroad in the last 3 years. 77\% participants confirmed that they did while 19\% didn't and 5\% cannot confirm.

These responses are significant because the aviation and maritime industry heavily depends on satellite communications and any major attack on satellite communication can disrupt the daily operation of these industries affecting millions of people including the participants of this survey.

Another very common usage of satellite is broadcasting television programs. According to the survey response, even today, 67\% of the participants or their neighbors have a satellite dish connected to their house for satellite TV. This resembles a big percentage of people still using satellite communications every day for entertainment and information purposes. Any adversarial impact (e.g., signal hijacking to spread propaganda which is described in detail later in this report) may include these audiences and millions of others throughout the world.

\begin{figure}
    \centering
    \begin{tikzpicture}
    \begin{axis} [
        ybar,
        bar width=15pt,
        ylabel=Number of Participants,
        xlabel=Impact Score,
        ymax = 13
    ]
    \addplot[
        color=orange,
        fill=orange
    ] coordinates {
        (1,2) 
        (2,2) 
        (3,1) 
        (4,2)
        (5,5)
        (6,6)
        (7,2)
        (8,11)
        (9,3)
        (10,9)
    };
    \legend {Average 6.91};
    \end{axis}
    \end{tikzpicture}
    \caption{Impact of attacks on Satellite Communications}
    \label{fig:impact_score}
\end{figure}
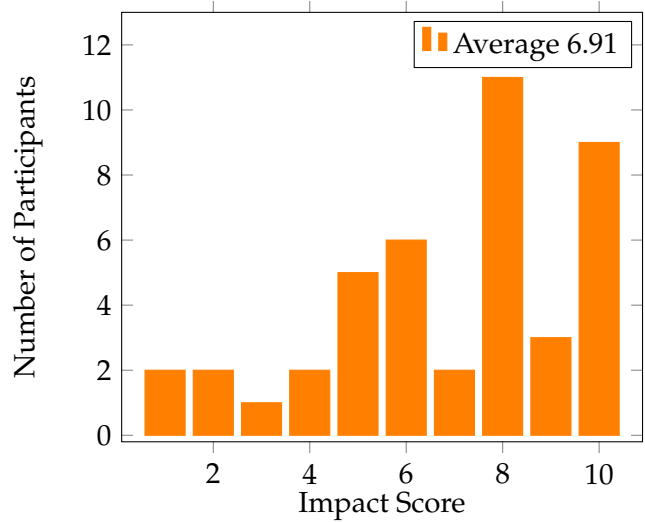

Then the survey asked the participants how severely they think they will be impacted if a major attack happens on the mainstream satellite communication network. They were asked to rate it on a scale of 1 to 10 where 1 meaning not impacted at all and 10 meaning severely impacted. The average rating recorded is 6.91 as presented in \ref{fig:impact_score}. This indicates that the general public is aware that they will be considerably impacted if a major attack happens on the mainstream satellite communication network.

\begin{figure}
    \centering
    \begin{tikzpicture}
    \begin{axis} [
        ybar,
        bar width=15pt,
        ylabel=Number of Participants,
        xlabel=Dependency Score,
        ymax = 11
    ]
    \addplot[
        color=cyan,
        fill=cyan
    ] coordinates {
        (1,1) 
        (2,0) 
        (3,3)
        (4,3)
        (5,4)
        (6,7)
        (7,4)
        (8,8)
        (9,4)
        (10,9)
    };
    \legend {Average 7.02};
    \end{axis}
    \end{tikzpicture}
    \caption{Dependency on Satellite Communications}
    \label{fig:dependency_score}
\end{figure}
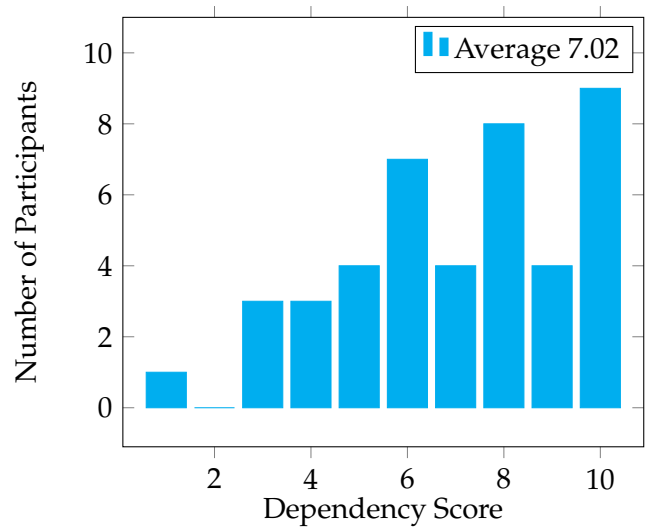

The survey also asked them how dependent they are on satellite communications in their understanding. Similarly, they were asked to rate it on a scale of 1 to 10. 1 meaning not dependent at all and 10 meaning severely dependent. The average recorded rating is 7.02 which again indicates that they are aware of their dependence on satellite communication for their everyday activities as presented in \ref{fig:dependency_score}.

Now that it is clear that satellite communications have significant impacts on the lives of people, the survey asked if they think "only" technological measures are enough to secure satellite communications. 16\% strongly disagree, 37\% disagree, 30\% were neutral and only 16\% agreed to this statement. This confirms that only technological controls are not enough to achieve resilience against satellite communication threats.

In the next question, the survey asked the audience if they think governance and compliance are required to secure satellite communication. Not a single participant disagreed with this. 60\% agreed and 40\% strongly agreed to this.

The survey asked if militaries, governments, and space agencies of different countries should work together to develop standards for satellite communication. 51\% strongly agreed, 37\% agreed, 9\% were neutral and only 2\% disagreed with this. From the outcome, this report concludes that all the aforementioned concerns should work together to come up with a robust standard framework for securing satellite communications.

\section{Key Takeaways}
The key takeaways from this section which will be considered while developing the proposed standard framework are stated below.

\keyword{Financially Impactful} -- Satellite communication is a financially impactful domain. Any adversarial activity can cause significant distress in both business and financial markets.

\keyword{UK Government Interest} -- The United Kingdom government is serious about the security of satellite communications and taking different initiatives regarding this. The author was privileged to be invited to one of the initiatives from the National Cyber Security Center of the UK. Moreover, people from the Ministry of Defense and National Cyber Security Center were interested in participating in the survey of this report.

\keyword{Considerable awareness} -- Majority of the audience understand that their everyday activities and business heavily depend on satellite communications.

\keyword{Beyond technology} -- Only technological solutions are not enough to secure satellite communications.

\keyword{Combined effort required} -- Governments, Militaries, and Space Agencies should work together to develop standards for governing satellite communications to achieve resilience.


\chapter{Satellites} 

\label{Chapter2} 


\setcounter{secnumdepth}{5}

\section{Elements and Segments of Satellite Communications}
The elements involved can be divided into three broad segments – Ground, Space, and the Ecosystem.

\subsection{Ground Segment}
The ground segment consists of the infrastructure on the earth. It can be on the soil, in the water (ocean or river), or even in the air (within a very close distance from the ground, e.g., airplanes). Based on the mobility, services at the ground segment can be divided into three broad categories - FSS, BSS, and MSS. Fixed satellite services are used for the stations that are located at fixed locations on Earth. Ground command and control stations for satellites, and fixed-location television stations are examples of FSS. When the earth stations (e.g., satellite signal receivers) are disbursed within a geo-diversified area and the same signal is required to be broadcasted to all of them, the service is used in BSS mode. When the receiving station is not fixed at a location and a specific unicast (mono or bi-directional) is required to be sent, the MSS is used.

\subsection{Space Segment}
The space segment consists of the satellites. A satellite is controlled with three basic subsystems – Telemetry, Tracking, and Command (TT\&C) (\cite{pelton_telemetry_2013}). TT\&C deals with the platform data. Platform data refers to the data or information related to the satellite itself. For example, the downlinked satellite's status, accurate location, uplinked control commands, etc. The TT\&C information is also used by the satellites themselves to maintain synchronization and communication between other satellites to establish the inter-satellite links (\cite{noauthor_ttc_nodate}).

\tocless\subsubsection{Types of Satellites}
Satellites are placed up in space, high above the Earth's ground and they orbit in different patterns. Based on the placement in the space, satellites can be categorized into 6 major categories.

\keyword{LEO} -- Low earth orbit satellites are closest to the earth. They are usually placed from 160 km to 1500 km above the ground. A LEO satellite can travel around the earth up to 16 times each day. Their orbital periods are usually from 90 minutes to 120 minutes.

\keyword{MEO} -- Medium earth orbit satellites are a bit further away. They are placed from 5,000 km to 20,000 km above the earth's surface level. Their orbital periods are between 2 hours to 12 hours. Because they are further away from the Earth, they can cover more surface on Earth than LEO satellites but there is propagation delay, and the signals are weaker than LEO satellites.

\keyword{GEO} -- Geosynchronous Equatorial Orbit or Geostationary orbit satellites have the highest coverage on the earth's surface because they are placed 35786 km above the ground and can cover 40 degrees of latitude on earth (\cite{peterson_satellite_2003}). They are placed over the earth's equator. Those are fixed on the same location in the sky.

\keyword{SSO} -- These satellites require calibration of the orbital inclination and altitude because the earth is not exactly round, and they cross specific locations precisely at the same local solar time. Their altitude is usually 600 km to 800 km.

\keyword{Lagrange Point Satellites} -- Lagrange points are very specific locations where the earth's gravity and the sun's gravity balance the orbital motion of the satellite (\cite{noauthor_nasa_nodate}). As a result, satellites in those points stay at the same place. Currently, there are only 5 Lagrange points identified by NASA which are L1, L2, L3, L4, and L5 (\cite{noauthor_what_nodate}). The recent James Webb Telescope is deployed on the L2 Lagrange point (\cite{noauthor_nasa_nodate}).

\begin{figure}
    \centering
        \includegraphics[scale=1]{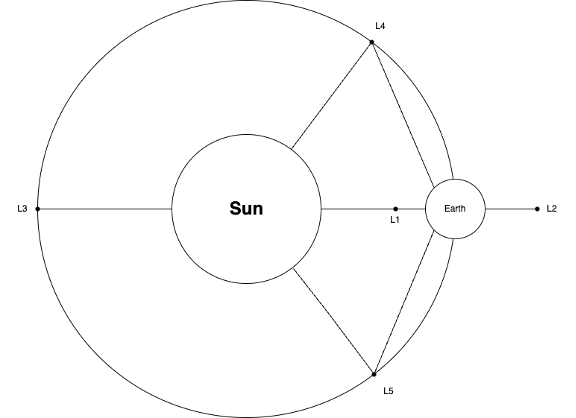}
        \caption{Lagrange Points}
        \label{fig:lagrange_points}
\end{figure}

\subsection{Ecosystem}
The microwave band has to be decided from the C-, X-, Ku-, Ka-, L- and S- bands. The orbit has to be decided and reserved. How the data will be sent and received (encryption and other cryptography mechanism) has to be decided. The payload design has to be architected. There are many more engineering aspects of the overall ecosystem that have to be decided to achieve effective and efficient satellite communication. Bruce R. Elbert has suggested a flowchart of the simplified steps involved in the implementation and operation of a satellite in his book (\cite{elbert_introduction_2008}).

\begin{figure}
    \centering
        \includegraphics[scale=1]{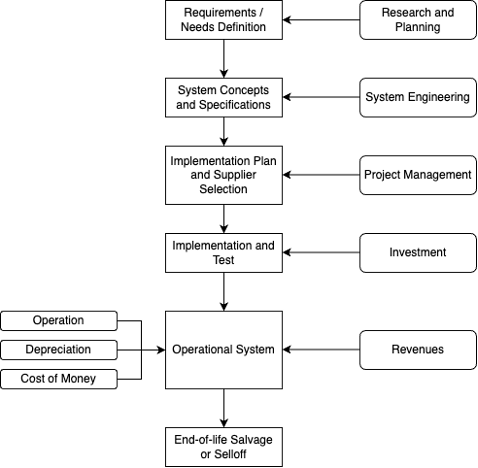}
        \caption{Satellite Implementation and Operation Steps. Source: (\cite{elbert_introduction_2008})}
        \label{fig:operational_steps}
\end{figure}

\section{History and Evolution of Satellite Communications}
The core technology behind satellite communication is microwave frequencies. The microwave frequency range used by satellites is from 1 GHz to 40 GHz (\cite{noauthor_satellite_nodate-2}). Microwave frequencies were not first used in satellites. Rather, they were used during World War II as a radar defense. Other common uses of microwave frequency include UHF mobile communication, VHF televisions, and HF radios.

Microwave frequency has to maintain a clear line of sight. If the path is long, then specialized repeaters have to be deployed to repeat the signal. Satellites can maintain a clear line of sight from the space. Due to the usage of microwave signals, satellites can be connected to and interoperated with terrestrial networks to a certain degree. The very early usage of satellite communication was transferring telephone calls from the gateway of one country to another. The gateway then would carry the signal forward through the terrestrial and other cable-based networks solving the problem of installing multiple repeaters in the path.

\begin{figure}
    \centering
        \includegraphics[scale=0.9]{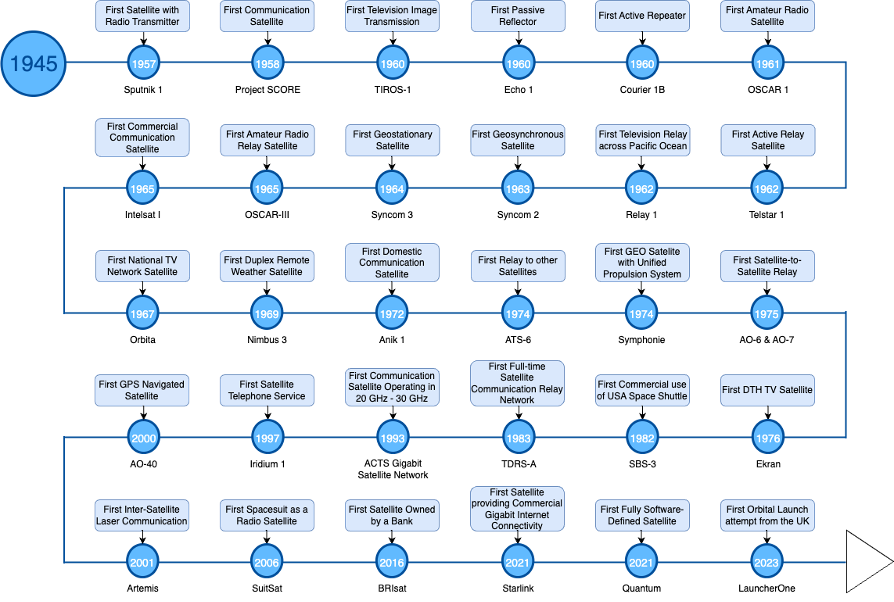}
        \caption{Timeline of First Significant Events in Satellite Communication}
        \label{fig:significant_events_timeline}
\end{figure}

\section{Lifetime Chart of Satellites}
\begin{figure}
    \centering
        \includegraphics[scale=1]{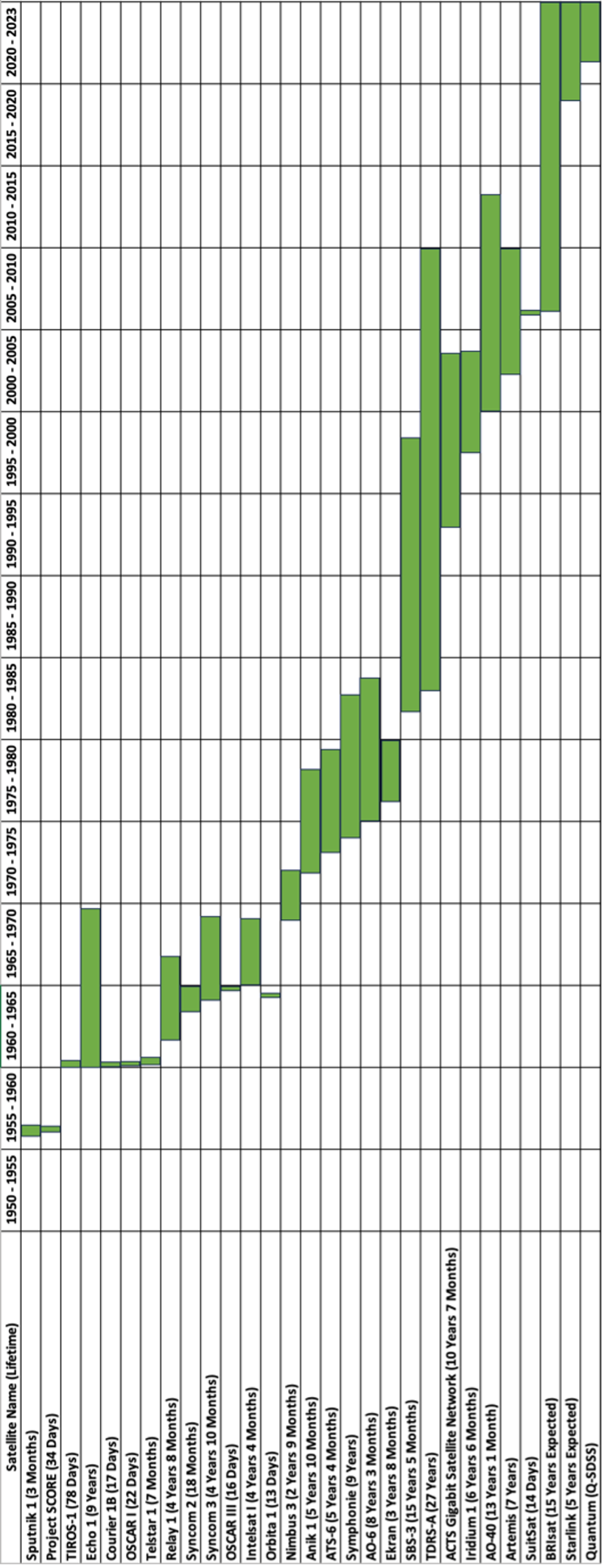}
        \caption{Lifespan of Initial Satellite Technologies}
        \label{fig:lifetime_chart}
\end{figure}

The lifetime of historically significant satellite missions is presented in \ref{fig:lifetime_chart}

\section{Key Takeaways}
The key takeaways from this section which will be considered while developing the proposed standard framework are stated below.

\keyword{Categorization criteria} -- Satellites can be generically categorized based on their distance from Earth.

\keyword{Impact of distance} -- Based on the distance, the availability of the attack vector varies. If a satellite is fixed in a position (GEO Satellites), then it is always exposed within the view of the adversary.

\keyword{Impact of position} -- If the satellite is not fixed but is part of a large constellation network (LEO Satellites), then it's exposed to attacks on the network and a compromised node can be utilized to attack other satellites in the network and beyond the network even though those are out of the view from a specific position on earth.

\keyword{Impact of lifespan} -- The longer the lifespan of a mission, the longer it needs to stay resilient against threats. This is challenging because this does not only consider the available threats during the deployment but also possible threat vectors that may arise in the future during the lifespan of a satellite.

\keyword{Attack launch segment} -- An attack can be launched at the ground segment, at the space segment, and even on the ecosystem of the satellite communication system.
 

\chapter{Attacks on Satellite Communications} 

\label{Chapter3} 


\setcounter{secnumdepth}{5}

\section{General}
Attacks on Satellite Communication do not only involve the cyber aspects but also the physical security of the satellite. This report will discuss the attacks on the physical layer of the satellite communication and attacks on the physical satellite to clearly differentiate between the attacks between the physical satellites and the communication of the satellite.

In 2007, a terrorist group hacked Intelsat satellite for spreading propaganda (\cite{bardin_satellite_2013}). In the same year, NASA Landsat-7 was also attacked by attackers and the satellite was unavailable for 12 minutes due to severe interference. On the next year, another NASA satellite Terra AM-1 got hacked twice causing multiple minutes of downtime in communication. Creech Air Force Base was attacked with malware on their satellite-based drone systems. In 2012, University of Texas researchers demonstrated a spoofing attack on satellite communication (\cite{bardin_satellite_2013}).

The cyber security of satellite communication depends on the infrastructure based on the requirements of the user. For example, a receiver receiving global position data will require more accuracy rather than the speed of data transfer. On the other hand, satellite internet will significantly focus on the data transfer rate preserving the common security services – confidentiality, integrity, and availability. Satellite phone calls may not need a high-speed data transfer rate but will still need security services. The physical security of the satellite mostly requires technology up in space. However, satellite communication security can be broken down into three areas – Ground Segment to Satellite Segment, Satellite to Satellite, and Satellite Segment to Ground Segment.

\section{Timeline of Attacks on Satellites}

\keyword{1963} -- The Soviet Union started the test of the co-orbital anti-satellite weapon in 1963. They made the system operational in 1973.

\keyword{1982} -- The United States of America started testing an advanced anti-satellite weapon – ALMV. The test was performed in 1985. The next few years were significant from the regulatory perspective, but all the major players in anti-satellite technology development understood that kinetic attacks are the most expensive and less precise. 

\keyword{2002} -- The United States deployed its first counter-communication system which can be controlled from the ground by jamming the target frequency.

\keyword{2005} -- NASA tested their first autonomous weapon that can collide with a target satellite without any instruction from the ground control systems.

\keyword{2006} -- China used ground-based laser technology to track satellites launched by the USA. The USA also started seeing an increase in jamming attacks in satellite communication.

\keyword{2007} -- China launched a weapon to test its capability by destroying one of its own old satellites. Also in 2007, the Sri Lankan rebel group "Tamil Tigers" was successful in hacking the Intelsat Satellite (\cite{noauthor_cyber_nodate}) to spread propaganda. Moreover, in 2007, NASA Landsat-7 got hacked and the communication was down for 12 minutes (\cite{vacca_cyber_2013}). 

\keyword{2008} -- The USA tried its first sea-based anti-satellite missile system. Also in 2008, another NASA satellite Tera AM-1 was hacked and was down twice in the same year for several minutes. 

\keyword{2010} -- India announced its development of laser-based anti-satellite weapons.

\keyword{2011} -- Iranian adversaries were successful in implementing malware in the Creech Air Force Base ground control station and were successful in spoofing GPS coordinates resulting in crashed drones (\cite{vacca_cyber_2013}).

\keyword{2012} -- A set of researchers from the University of Texas, Austin could replicate a similar spoofing mechanism in satellite communication.

\keyword{2014} -- Chinese adversaries were successful in disabling communication with a USA weather satellite (\cite{noauthor_cyber_nodate}).

\keyword{2022} -- Russian hackers were accused of jamming internet communication of the Starlink satellite network. Moreover, in the same year (2022), a wiper malware-based cyberattack was reported on the KA-SAT satellites (\cite{noauthor_viasat_nodate}). The malware name is "AcidRain", and it wiped out satellite internet modems. Viasat had to redeploy nearly 30,000 satellite modems to restore internet communication after this malware-based cyberattack on satellite communication. 

\keyword{2023} -- Russian military satellite communication was disrupted by a cyberattack (\cite{noauthor_cyberattack_nodate}).

\begin{figure}
    \centering
        \includegraphics[scale=0.9]{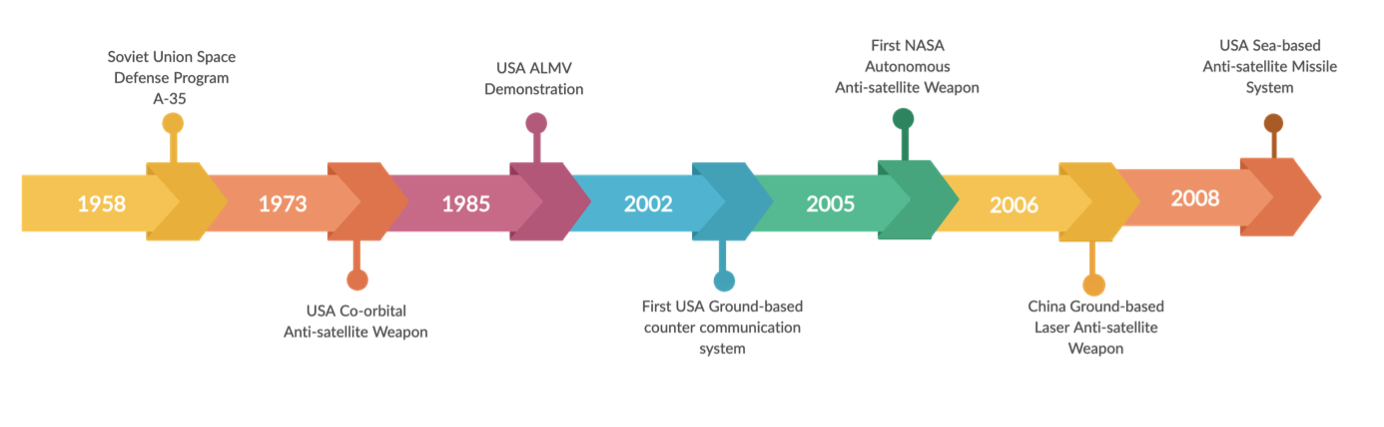}
        \caption{Anti-satellite Development Timeline}
        \label{fig:attack_timeline}
\end{figure}

\section{Attack Vectors for Satellite Communications}
\subsection{Physical Attacks on Satellites}
\tocless\subsubsection{Kinetic Energy Attacks}

Common Orbit Anti-Satellite Interceptors are placed in the same orbital height of the satellite. Once the command is sent from the ground segment, they approach the target satellite through orbital maneuver and apparently collide with the target satellite and either destroy or remove the satellite from the expected position to interrupt the satellite communication. The greater risk associated with this type of weapon is – it can be latent for a long time and be activated when required. They can reach their target within a few hours of time from activation.

The more advanced version of this weapon is called Active Collision Attacker. They can set their own maneuver autonomously and reach the target and destroy the target satellite by colliding. An example of such a weapon is the "Deep Impact" spacecraft launched by the USA (\cite{noauthor_mission_nodate}). This spacecraft was launched to impact a comet – Tempel 1. However, it is possible to use a similar spacecraft/weapon to attack another satellite.
Based on the type of impact, kinetic energy attacks can be divided into three categories:

\keyword{Direct Ascent Anti-satellite Weapon} -- This is also known as kinetic-kill attack (\cite{noauthor_ground-based_nodate}). In this attack, the kill object, also known as a projectile, is directly launched toward a target satellite to damage or destroy it with kinetic energy. This can be launched from air, sea, or even from aircraft.

\keyword{Pellet Cloud Anti-satellite Weapon} -- In this attack, the weapon does not directly impact the satellite itself. Rather, it releases a cloud of densely packed elements right in the path of the satellite coverage to disrupt satellite communication.

\keyword{Explosive Anti-satellite Weapon} -- In this attack, instead of colliding directly with the target satellite, a warhead is detonated near the target satellite, so the satellite is damaged by the debris. Because the direction of the debris cannot be controlled, this is less effective compared to the other two types of ASAT weapons.

Based on the target of the impact, these warheads have some specific requirements to be successful as presented in \ref{tab:capabilities_required}

\begin{table}
\centering
\begin{tabular}{l l l l}
\toprule
\tabhead{Capability} & \tabhead{Direct ascent} & \tabhead{Pellet cloud} & \tabhead{Explosive} \\
\midrule
Suborbital launch & Yes & Yes & Yes\\
Orbital launch & Yes & Yes & Yes\\
Manoeuvrability & Yes & No & No\\
Precision Tracking & Yes & Yes & No\\
Approximate Tracking & No & No & Yes\\
Autonomous Tracking & Yes & No & No\\
\bottomrule\\
\end{tabular}
\caption{Capabilities required by ground-based ASAT (\cite{noauthor_ground-based_nodate})}
\label{tab:capabilities_required}
\end{table}

\tocless\subsubsection{Direct Energy Attacks}
The main concept behind this type of attack is focusing and beaming high energy in a certain direction toward the target satellite to damage the components of the satellite. Unlike sending a warhead to space to impact the satellite, this type of attack can be done within a very short time as the beam travels nearly at the speed of light. Zhenhua Liu et al. (\cite{liu_space_2020}) identified this as the main research direction after the kinetic energy satellite attacks. Up until now, three types of beaming have been developed to attack satellites.

The first type is using high-energy laser beams. It was calculated through simulation that a 3 MW ground-based laser beam can generate 250 W/cm2 energy density at a distance of 300 kilometers which can be lethal and cause unrepairable damage to the photoelectric sensors, satellite solar cells, and/or the satellite thermal control system. However, the energy density becomes lower at further distances. The simulation shows that at a 1000-kilometer distance, the energy density becomes only 25 W/cm2 but this still can cause significant damage to the sensitive satellite components. The temperature in the space is usually extreme. For example, the sun-facing side of the International Space Station can reach up to 121°C but the dark side can go down up to -157°C. Due to this extreme condition, satellite thermal control has to be effective and functional for optimum satellite communication. However, any adversarial energy can cause exceeding of the highest tolerable temperature of the components of the satellite and thus damage the components. Currently, the USA holds ground-based weapons that can emit laser beams with 10 MW of power (\cite{liu_space_2020}).

The second type utilizes high-energy particle beams. In this approach, particles like electrons, protons, and ions are accelerated and gathered into a dense beam. Then the beam is pointed to the target satellite. Particle beam can travel near the speed of light, so it provides a very fast attack timing. This type of weapon can be divided into two major types – charged particle beam, and neutral particle beam. However, this is still in the research stage, and up until now, there has not been any practical application of this type of weapon.

Another important type of such weapon can be the high-energy microwave. In this approach, a high-energy microwave beam is pointed and transmitted towards the target satellite. The concentrated energy in the microwave can destroy the target satellite easily. According to Zhenhua Liu et al. (\cite{liu_space_2020}), the key technology in high-energy microwave weapons is the pulse power source technology, antenna technology, ultra-short pulse technology, and ultra-high bandwidth. This is also known as HPM. This technology started its journey back in 1973. Back in 2005, the USA Air Force proposed putting HPM generator devices into the satellites to attack other satellites (\cite{liu_space_2020}).

\subsection{Cyber Attacks on Satellite Communications}
The cyber threats to satellite communication, in general, is the particular area of interest of this report because this is the area that deals with the cyber elements of satellite communication. It can be categorized from various perspectives. For example, attacks can be targeted on basic security services like Confidentiality, Integrity, and Availability because these are the fundamental goals of securing satellite communication.

\tocless\subsubsection{Cryptographic Attacks}
\begin{figure}[ht]
    \centering
        \includegraphics[scale=0.4]{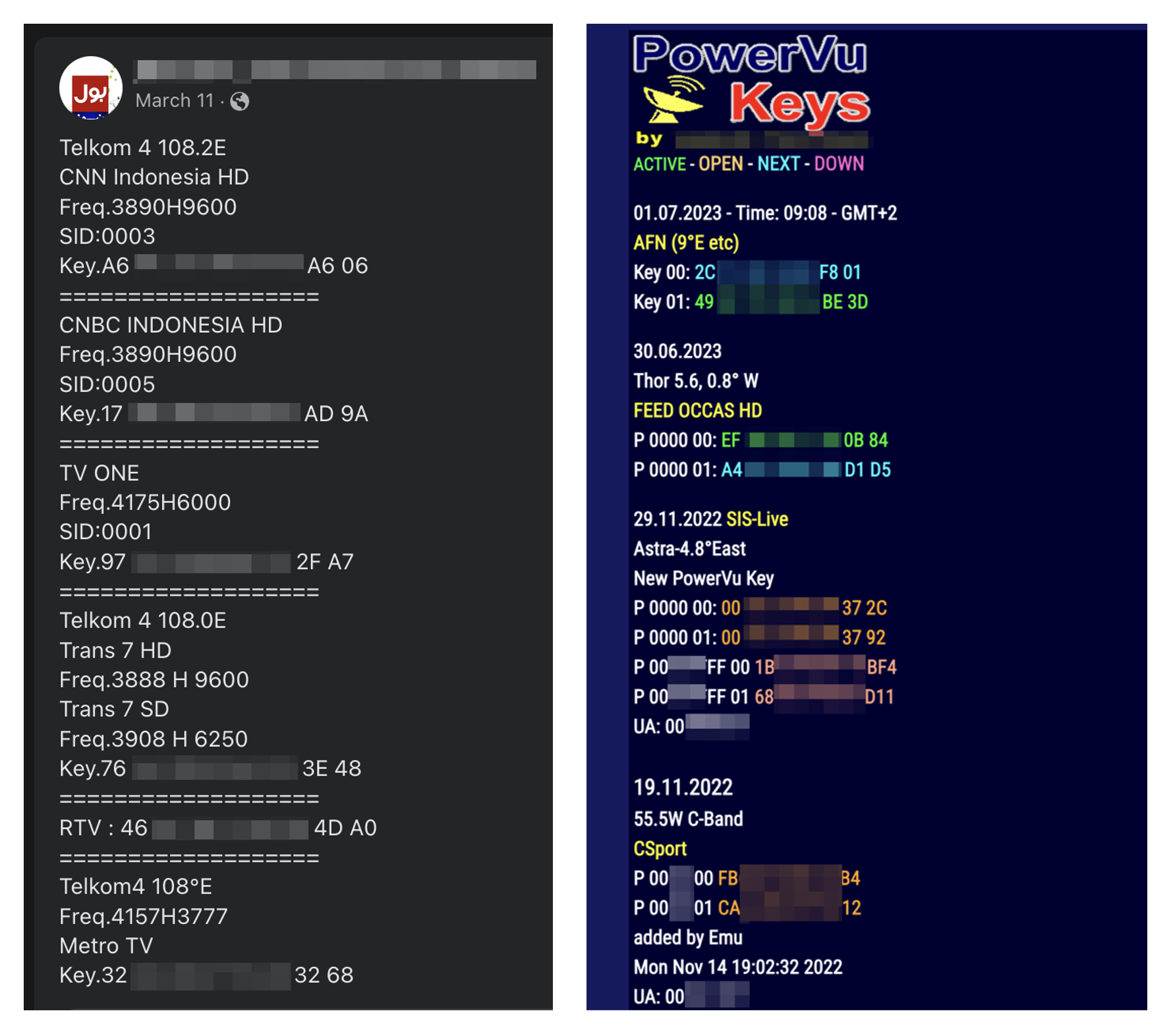}
        \caption{Unauthorized Publicly Disclosed BISS and PowerVu Key Example}
        \label{fig:cryptographic_attacks}
\end{figure}

To understand the cryptographic threats and attacks better, we need to focus on the common cryptographic technologies used today for securing satellite communication. There are various proprietary and public technologies available today for encrypting the information transferred over the satellites. One of the most common technologies for scrambling data from unauthorized receivers is – BISS (\cite{noauthor_biss2_nodate}). The drawback with this method is – it can be attacked with brute force easily and the key can be retrieved within a reasonable time with regular household computing power (\cite{noauthor_how_nodate}). It requires software named "CW List Brute Force" or "CWBruteList" to brute-force and find the correct BISS key.  Moreover, a lot of adversaries unauthorizedly share the BISS key publicly for many satellite TV channel streams and those keys can easily be obtained from the internet. This threat utilizes a generic cyber-security unauthorized disclosure risk to attack the confidentiality of satellite communication as demonstrated in \ref{fig:cryptographic_attacks}.

Another common method for protecting communication is PowerVu. This is used by many popular broadcasting service providers as of today. However, just like the BISS scrambling system, the decryption keys are also leaked by the attackers publicly in a similar unauthorized manner which is demonstrated in \ref{fig:cryptographic_attacks}.

In North America, the DigiCipher signal transmission standard is also quite popular. This is a proprietary standard for broadcasting video over satellite communication. In this technology, a virtual channel number is used which is only known by the authorized receiver. The virtual channel number is used along with the satellite's downlink frequency to secure the communication. This is more secure than the previously discussed BISS and PowerVu technology. However, there is already a proof of concept available for unauthorizedly decrypting DigiCipher II (the latest iteration of the DigiCipher standard) (\cite{gerlinsky_how_nodate}).

\tocless\subsubsection{Eavesdropping}
Satellite signals are broadcast in general. That means, any party with proper information and relatively cheaper equipment can receive the signals. If the signals are unencrypted, then any party, regardless of whether authorized or not, can receive the original message. Eavesdropping is not only used for extracting information, but also for gathering intelligence about the communication that can be later used to develop further attacks.

\tocless\subsubsection{Software Threats}
Programming and architectural level vulnerabilities may be exploited by the attackers utilizing the Microarchitectural Side Channel Attacks to extract information unauthorizedly. These attacks can be very complex and can be designed based on timing information and performance counters. Moreover, memory-based attacks are theoretically possible as presented by previous research (\cite{cassano_is_2022}).

\tocless\subsubsection{Hardware Threats}
Maliciously or poorly architected hardware has the possibility to be exploited by software, especially in the case of hardware trojans. Moreover, due to the appearance of new vulnerabilities in the hardware design or processors, relatively safer hardware may become insecure once a vulnerability is discovered. Due to the distance from Earth and the speed of travel, it's immensely difficult and costly to replace any hardware on a satellite that is already placed in orbit. Due to limitations of the architecture and design, sometimes, components used in satellites may have weaknesses that can be exploited by attackers to unauthorizedly access information in satellite communication. Previous research has discussed security risks associated with a common RISC-V microprocessor used in satellites (\cite{cassano_is_2022}).

\tocless\subsubsection{Unauthorized Use of Transponder}
Satellites with transparent forwarding functionality and with no onboard signal processing capability are vulnerable to unauthorized use of the transponder. Because those satellites do not process the signals and simply forward them by reflecting, an adversary with proper information about the unused channels can utilize this loophole and unauthorizedly use the expensive transponders. There have been records of this illegal communication available on social media platforms where researchers could eavesdrop on the illegal communication of the pirate using the USA Navy Satellite Transponders unauthorizedly (\cite{noauthor_pirates_nodate}). Very little information, (for example, the satellite orbit information, and operating frequency) is required to unauthorizedly use the transponders of the satellites. Sometimes, the attackers use unused channels. However, more smart attackers can utilize the Direct Sequence Spread Spectrum and Power Spectral Density technique to hide the communication within the currently used frequency band of the satellites. Frequency hopping is a pragmatic solution to this problem. However, frequency hopping itself alone cannot solve this problem totally.

\tocless\subsubsection{Spoofing}
Spoofing is an advanced technical attack on satellite communication that involves sending malicious signals in the disguise of original signals. This is advanced because, in this method, the attacker has to have advanced knowledge of the original signal (for example, physical layer waveform, signal characteristics, frame structure, etc.) so the signal can be forged in such a way that the receiver device can be deceived, and the receiver device will accept the malicious signal as the original one. An example of satellite signal spoofing can be GPS spoofing. According to previous research (\cite{he_communication_2016}), GPS signals are comparatively easier to spoof as the signal structure and other parameters are known.

\tocless\subsubsection{Unauthorized Modification of Information}
Sometimes, instead of spoofing, attackers can intercept the signal, modify it unauthorizedly, and then rebroadcast it. If the transmitted information doesn't have any mechanism involved to protect the integrity of the information or doesn't have data origin authentication, then this type of attack cannot be detected and stopped.

\tocless\subsubsection{Denial of Service}
This is one of the common targets of the adversaries. In this type of attack, the adversaries implement various methods and techniques to make the satellite communication unavailable to the target receivers. Previous research has identified different methods for this attack (\cite{yue_low_2023}). However, one of the common approaches is flooding the satellite with continuous requests. If the satellite doesn't have a mechanism deployed to detect anomalies in requests or cannot identify if the request is coming from legitimate sources, then this attack can be fruitful for the adversaries. As the transponders get busy dealing with unauthorized requests, they cannot respond to the authorized requests anymore. As a result, legitimate users cannot get the desired communication service from the satellite. Jamming is another technique to achieve the Denial of Service from satellites.

\tocless\subsubsection{Node Compromise}
Another potential attack can be done by compromised nodes in the satellite constellation (\cite{roy-chowdhury_security_2005-1}). If a node can be compromised by exploiting any vulnerability in the hardware or software or in the command-and-control communication system, then it can be used for adversarial purposes. This is a critical risk because once a node is compromised, the receivers or other satellites cannot easily distinguish the signals coming from a compromised node. The signals from a compromised node can look absolutely normal to other satellites and the receivers because the signals are indeed coming from the satellite itself. The receiver ends have to have sophisticated technology to detect anomalies in signal and data patterns to suspect any issues in the communication signals. However, it is comparatively easier for the ground control stations as they have full access to the satellite. If they have a proper detection mechanism installed, then they can identify that a node is compromised faster than any other parties. This can result in sending malicious signals to compromise other nodes in the constellation. If the command-and-control communication is compromised, it can be driven toward other satellites for direct collision. Apart from this physical attack through a cyber technique, a compromised node can be used for various other non-physical adversarial purposes too. For example, it can be used to spread malicious and incorrect information to damage/destroy other critical systems depending on satellite communication. For example, wrong coordinates or global positions can cause significant damage to a vessel in the ocean or an aircraft up in the sky. Besides, sensitive information relayed through the satellites can be compromised from a compromised node. On top of that, it can also leak significant sensitive information about the satellite itself or the organization controlling the satellite that can help in developing future attacks on the satellites from the same or similar organization.

\tocless\subsubsection{Power Based Jamming}
Signal jamming is a common method for attacking satellite communication. These attacks can be divided into two major areas – Suppressive Interference and Deceptive Interference. Suppressive interference involves sending malicious signals either in the same frequency as the target satellite communication or within the band of the target satellite communication. On the other hand, deceptive interference involves transmitting identical but incorrect signals to the receiver to confuse them and interrupt the communication. This type of attack is fairly easier compared to other complex modes of attack. To make this kind of attack, adversaries use power-based jamming signals to disrupt the original signal from the satellite. There are several types of power-based jamming techniques as discussed in previous research (\cite{zou_survey_2016}) that can significantly disrupt satellite signals as the signals are airborne and there is a significantly large distance between the sender, the satellite, and the receiver. This kind of attack can be divided into three areas.

\keyword{Ground-based jamming} -- The ground-based jamming consists of jamming attacks both from the soil (fixed location or moving vehicle with jamming equipment) and the water (jamming equipment on boats or ships). Ground-based jamming is usually more powerful because of the prolonged visibility of the satellites. It also involves less cost and has access to more equipment ready for this kind of attack. Ground-based jamming usually focuses on blocking the uplink signal by sending jamming signals. The main aim of ground-based jamming is to block access to the actual transponder on the satellite. Sometimes this type of attack is used to transmit a modified malicious signal to the target receivers (\cite{cassano_is_2022}).

\keyword{Air-based jamming} -- A Similar type of attack can also be launched from aircraft. Air-based jamming attacks can have an impact on both uplink and downlink as the aircraft fly above the ground and have more coverage than ground attack terminals.

\keyword{Space-based jamming} -- A jamming attack can also be launched from space – from other satellites. However, due to the orbiting nature of the satellites, these attacks are less powerful than the previous two because of the coverage. Space-based jamming attacks mostly have an impact on the downlink but cannot be very effective for jamming uplinks. This is because spacecraft don't usually have the high energy source required to have an impact on a ground station.

Based on the target segment (ground or space), jamming attacks can be classified into two types.

\keyword{Orbital Jamming} -- This can also be referred to as uplink jamming. In this jamming attack, powered signal beams are thrown toward the satellite itself to interrupt the legitimate transmission capabilities of a satellite. In this attack, the target is sending malicious and rigorous continuous uplink signals to the satellite to overwhelm the transponders of the satellite. In most cases, this results in Denial of Service.

\keyword{Terrestrial Jamming} -- This is similar to orbital jamming, but the difference is – in this attack, the ground receivers are targeted. This can also be referred to as downlink jamming. In this attack, malicious signals are sent to the receiver antennas to distort the received signals. This also, in most cases, results in Denial of Service.

\section{Timeline of Significant Satellite Jamming Capabilities Development in Last Two Decades}
\keyword{2004} -- USA demonstrated capabilities of interfering satellite communication with radiofrequency.

\keyword{2010} -- Russia demonstrated capabilities for suppressing communication channels using radio signals using their Borisoglebsk-2 system (\cite{noauthor_electronic_nodate}).

\keyword{2015} -- China demonstrated capabilities to interfere and jam satellite, radar, and GNSS communications systems (\cite{noauthor_ground-based_nodate}).

\keyword{2018} -- Russia demonstrated signal jamming capabilities in the range of 300 MHz to 3 GHz with their Tirada-2 and Bylina-MM systems (\cite{noauthor_space_nodate-1}).

\keyword{2020} -- USA released an updated version of their satellite communication interfering capability (Block 10.2).

\section{Key takeaways}
The key takeaways from this section which will be considered while developing the proposed standard framework are stated below.

\keyword{Significant Threats} -- Two significant threats for satellite communications are physical attacks and cyber-attacks.

\keyword{Physical attack} -- Physical attack mitigation is beyond the scope of this report.

\keyword{Common attacks on wireless} -- A reasonable type of common cyber-attacks on wireless and other systems are impactful for satellite communications.

\keyword{Frequent power-based jamming} -- According to the incident reports in the public media, power-based jamming attacks have significantly increased over the last two decades.

\keyword{Single control is insufficient} -- No single mitigation technology and security control is enough to provide overall resilience against cyber-attacks on satellite communication.

\keyword{Effective approach} -- The most effective approach will be developing a robust satellite communication security framework that can be adhered to by the stakeholders and implemented according to the requirements of each specific space mission.


\chapter{Analysis of Existing Standards for Secure Satellite Communications} 

\label{Chapter4} 


\setcounter{secnumdepth}{5}

\section{General}
All space agencies and private corporations have their standard, policies, and procedures. Some are public, some are proprietary. This project only looked at the publicly published security standard documents for secure satellite communication which are significantly related to the cyber security of the satellite communication. The list is demonstrated in table \ref{tab:standard_list}.

\begin{table}
\centering
\begin{tabular}{l l l l}
\toprule
\tabhead{Topic} & \tabhead{Title} \\
\midrule
Information Assurance for Space\\ Systems to support National\\ Security Missions & CNSSI 1200 (\cite{noauthor_national_2014})\\
\midrule
Space Platform Overlay & CNSSI 1253F Attachment 2 (\cite{noauthor_security_2014})\\
\midrule
Cybersecurity for Commercial\\ Satellite Operations & NIST IR 8270\\
\midrule
Cybersecurity Framework to\\ Satellite Command and Control & NIST IR 8401\\
\midrule
Security Architecture for\\ Space Data Systems & CCSDS 351.0-M-1 / BS ISO 20214:2015\\
\midrule
Cryptographic Algorithms for\\ Space Data Systems & CCSDS 352.0-B-2 / BS ISO 21324:2016\\
\midrule
Space Data Link Security\\ Protocol & CCSDS 355.0-P-1.1\\
\midrule
Network Layer Security\\ Profiles for Space Data Systems & CCSDS 356.0-B-1\\
\bottomrule\\
\end{tabular}
\caption{List of public standard documents for secure satellite communication}
\label{tab:standard_list}
\end{table}

\section{Analysis of the identified standards}
If we look at the first document in Figure 14 which is CNSSI 1200, it completely focuses on national security systems deployed in the space for the USA. Section I and Section III, clarify that anything that does not deal with the US national security systems will be beyond the scope of the document. This indicates that it will not put any recommendation related to satellite communication for other purposes of satellite communication including earth and space observatory satellites, public communication satellites, or navigational satellites which hold a significant portion of the satellite usage today. Due to the intrinsic nature of this standard, we cannot consider this as a generic baseline standard for the overall security of satellite communications. However, we have considered the pragmatic guidelines suggested in CNSSI 1200 that can help come up with a generic standard of guidelines for securing satellite communications regardless of usage.

Another significant standard document from the CNSSI is the CNSSI 1253F Attachment 2 (\cite{quiquet_cnssi_nodate}) which is an overlay standard that provides guidelines only for the unmanned spacecraft in the space segment for the national security systems for the USA. It specifies in Section 1 that it does not provide any recommendation for the ground segment or user segment. Again, a generic standard should provide guidelines on all common areas of satellite communication and that is why this report has considered the guidelines for the space segment that can be helpful for all segments of satellite communication while developing a universal standard for secure satellite communication.

The next utterly important standard document we considered is the NIST IR 8270. This is important from quite a few different characteristics. First of all, this was published in July 2023 and is the most recent standard regarding cybersecurity for satellite communications. It contains the most recent viewpoints and guidelines from the USA in the cybersecurity domain (which is the area of concern of this report) for satellite communications in commercial satellite operations. Commercial use of satellites holds the largest portion in the usage segment of satellites. This is the second important characteristic of this standard document. The third significant characteristic of this standard is that – it recognized space as the commercially critical infrastructure. Commercially critical infrastructures play a vital role in many different sectors of business and finances of the whole world. Because of these characteristics, we will critically analyze this standard and will accommodate the relevant recommendations from this to our proposed standard for satellite communication security.

NIST IR 8401 is an important standard document however its scope is limited to the ground segment of the satellite communication only. It provides useful guidelines for the command-and-control section of the ground segment in satellite communication. Besides, this is very useful for organizations in risk management for other elements like systems, networks, and assets in the ground segment of satellite communication. This report will carefully analyze and welcome the effective guidelines from NIST IR 8401 while developing the proposed universal satellite communication security guideline.

The next 4 standards which were taken into consideration are from the CCSDS. These standards are significantly important for the proposed solution for several purposes. First of all, the BSI has accommodated two of the standards as it is in the ISO standard documents. Secondly, CCSDS is a forum of major state leagues in the satellite development, deployment, and operations industry. The agencies are listed below.

\begin{itemize}
    \item Agenzia Spaziale Italiana (ASI)
    \item Canadian Space Agency (CSA)	
    \item Centre National d'Etudes Spatiales (CNES)	
    \item China National Space Administration (CNSA)	
    \item Deutsche Zentrum für Luft- und Raumfahrt (DLR)	
    \item European Space Agency (ESA)	
    \item Instituto Nacional de Pesquisas Espaciais (INPE)	
    \item Japan Aerospace Exploration Agency (JAXA)	
    \item National Aeronautics and Space Administration (NASA)	
    \item State Space Corporation (ROSCOSMOS)	
    \item United Kingdom Space Agency (UKSA)
\end{itemize}

Because of this collaborative nature, the CCSDS standard documents are more robust and inclusive than any other standards that were analyzed in this report. Due to this distinctive characteristic, CCSDS standards will be used as the baseline template for the proposed standard in this report. However, there are significant areas in the main standard document, CCSDS 351.0-M-1 which is represented as it is in the BS ISO 20214:2015, that does not cover the important advancement in the understanding and requirements of the current cybersecurity landscape. For example, the CCSDS 351.0-M-1/BS ISO 20214:2015 guides inherent trust in the participating parties in the development and operations of satellites while the modern cyber security architecture strongly recommends a zero-trust model (\cite{noauthor_zero_nodate}). The CCSDS document that was incorporated into the BS ISO 20214:2015 was published in November 2012 which is nearly a decade old now today. Due to the nature of rapid progress in the threat landscape, it requires a pragmatic upgrade incorporating the latest security architecture and philosophy. That is the main goal of this report – taking the previous standard as a baseline and proposing an effective standard for secure satellite communication architecture.

\section{Key takeaways}
The key takeaways from this section which will be considered while developing the proposed standard framework are stated below.

\keyword{Private standards} -- Each space agency (including, but not limited to, militaries, governments, and private agencies) has its strategic policies for securing its space systems.

\keyword{US standards} -- The USA has considerably robust security frameworks for their space missions compared to any other country.

\keyword{Considerable standards} -- CCSDS forum standard documents are the most considerable satellite security standards for the United Kingdom as of today.

\keyword{No UK-focused standard} -- There is no specific public cybersecurity standard framework that focuses on the interest of the United Kingdom and its ally nation states solely.

\keyword{11 years old standard} -- The most relevant framework is the BS ISO 20214:2015 which has not been updated in the last 8 years. The BS ISO 20214:2015 simply put the CCSDS 351.0-M-1 as-it-is inside the ISO standard document. Surprisingly, the CCSDS 351.0-M-1 was last updated in 2012, which is 11 years old now.

\keyword{Standard update required} -- This confirms that there is a large opportunity and need for an updated and pragmatic generic security standard for satellite communication and space data systems that will serve the interest of the United Kingdom and its ally state nations with the top priority and also help other space agencies to use this as the baseline while designing their space mission systems.
 

\chapter{Enterprise Architecture Frameworks Review and Analysis} 

\label{Chapter5} 


\setcounter{secnumdepth}{5}

\section{General}
Enterprise information system architectures provide a very high-level blueprint for any organization. These provide a holistic approach to system design, planning, and implementation. The main goal of enterprise architectures is to guide organizations to achieve their business goals and translate the business strategy to information technology for achieving the goals. That is why, enterprise architectures will be considered out of scope for this proposed secure satellite communication framework. However, this report will analyze the popular and widely accepted enterprise architectures and security-specific architectures and models to find out if those can be considered for the overall cybersecurity of satellite communications and space data systems.

\section{Enterprise Architectures}
\keyword{SABSA (Framework)} -- A methodological and business-driven security framework. It does not provide any specific security control. Rather, the framework provides six layers of architecture having business goals and requirements at the top to ensure business alignment. Because of the architectures of this framework, this is well aligned with operations which are mostly business-focused within a controlled and limited scope. However, because of the high diversity in satellite communication architecture, the SABSA framework does not seem to be the best approach to integrate within satellite communication standards.

\keyword{Zachman (Framework)} -- A very popular enterprise system. It was published in 1987 in the IBM system journal (\cite{zachman_framework_1987}) and since then it has gained significant traction in research in academia and in the industry. This framework provides a strong overview of enterprise information systems from different viewpoints. It also provides a clearer picture of the interconnections between the components within a system.

Even though this is a great template framework, there are considerable flaws in the analogy of this framework. Even the colleagues of John Zachman (who published this framework) discovered flaws in the analogy explained by John Zachman in his Zachman Framework (\cite{stecher_building_1993}). There have been reports of organizations facing adverse and negative impacts instead of the assumed theoretical benefits following the Zachman framework (\cite{ahlemann_strategic_2012}). The crucial requirements of the Zachman Framework have not proved to be fruitful. However, it is accepted that the framework worked as a template framework for many other effective frameworks widely implemented in the industry in the past few decades.

\keyword{TOGAF (Framework)} -- The full form of this framework is The Open Group Architectural Framework which was created in 1995 and owned by The Open Group. This is purely based on the USA Department of Defense Architecture Framework (DoDAF). This is one of the most heavily utilized enterprise architectures in the industry. TOGAF Frameworks consists of 6 core elements. The ADM provides guidelines with a sequence of different phases and steps for developing the architecture. However, following the specific sequence and procedures will narrow down the path while developing the space data systems and satellite communications. Satellite Communication and Space Data Systems are widely diversified in nature and architecture. That's why TOGAF can include obstacles without many benefits for the area of interest of this report.

\keyword{DoDAF (Framework)} -- DoDAF's first version was released in August 2003. The full form is – the US Department of Defense Architecture Framework. As it was developed by the US Department of Defense, it mainly focuses on defense systems. DoDAF provides specific guidelines for the architecture and the products to clearly describe the system and specific steps to develop the deliverables. Again, because of the widely diversified requirements and deliverables from space data systems, this may or may not be the best architecture to propose for satellite communications and space data systems.

\keyword{MODAF (Framework)} -- MODAF – UK Ministry of Defense Architecture Framework is similar to DoDAF in the sense that it is developed by the UK counterpart of the team developing DoDAF. MODAF used DoDAF version 1 as the baseline while developing it. MODAF provides views in 7 categories that represent and give a clearer picture of the whole business. This is helpful in understanding a complex issue within the business process. It is useful to acquire and represent data in a rigorous, coherent, and comprehensive method that enables it to facilitate an easier understanding of the complex problem and solve it. The advantage of MODAF is that it focuses more on the business perspective rather than the information technology. This particular advantage is a disadvantage for the interest of this report.

\keyword{NAF (Framework)} -- The next consideration of this report was the NATO Architecture Framework. It is a great framework that provides useful guidance to organize and represent the architecture to the stakeholders through its two-dimensional grid scheme. It was released in 2018. It provides guidance on both system architectures and enterprise architectures. It is basically a content-focused architecture framework. Unlike DoDAF, it doesn't guide with any method or process. The strongest point about NAF is that it supports interoperability between the frameworks of multiple state nations. However, this doesn't particularly focus on the security of the information technology. That is why this report can't heavily consider this while proposing the security framework for satellite communications.

\section{Security Domain Models}
\begin{figure}[ht]
    \centering
        \includegraphics[scale=0.6]{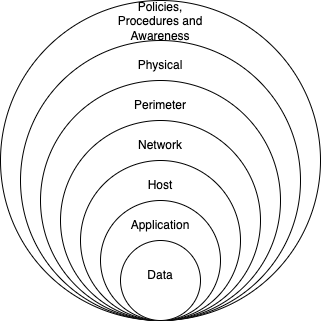}
        \caption{Defense in Depth Model}
        \label{fig:defense_in_depth}
\end{figure}

\keyword{Defense in Depth (DiD) Model} -- DiD model is particularly interesting for this report because this cybersecurity model provides multi-layer defensive mechanisms to protect data and information as represented in \ref{fig:defense_in_depth}. One of the core beauties of this model is that it supports redundancy which makes it difficult for the adversaries to achieve desired damage in terms of security. However, due to the nature of this model, it is relatively costly to implement, and the architecture of satellite communications and space data systems does not have much scope for redundancy, at least in the space segment.

\keyword{Zero-Trust Model} -- According to Microsoft (\cite{mjcaparas_what_nodate}), Zero-Trust is not exactly a security framework or a model, rather it is a security strategy. Microsoft identified 3 principles as the guiding principles of Zero-Trust. Microsoft recommends that different information technology areas be made more secure with Zero-Trust including secure identity, secure endpoints, secure applications, secure data, secure infrastructure, secure networks, secure visibility, and automation. The core beauty of the Zero-Trust strategy is that it does not trust any entity by default, rather, it trusts by exception with explicit verification. US NIST Framework considered Zero-Trust as a set of evolving cybersecurity paradigms (\cite{rose_zero_2020}).

The effectiveness and robustness of zero-trust architecture are so accepted that the US released an "executive order 14028" (\cite{house_executive_nodate}) that directed federal agencies to utilize the zero-trust strategy to secure federal government cybersecurity.

The World Economic Forum also recognizes Zero-Trust as an effective security framework (\cite{noauthor_zero_nodate-1}) as seen in the community paper published in August 2022. 

Because of these distinctive features and absolute focus on information security and the capability to confine any attack and damage, this will be the most appropriate and effective strategy to integrate into the satellite communication and space data systems. This report will heavily consider the zero-trust framework strategy while developing the proposed update of the satellite communication standard.

\section{Key takeaways}
The key takeaways from this section which will be considered while developing the proposed standard framework are stated below.

\keyword{Too broad scope} -- Enterprise Architectures are too broad in scope and mostly focus on the business instead of information security.

\keyword{Aligned strategies} -- Security domain models are more aligned with the scope of this report and this report will heavily utilize the benefits of the Zero-Trust model within the proposed guidelines.


\chapter{Secure Satellite Communications Standard Proposal} 

\label{Chapter6} 


\setcounter{secnumdepth}{5}

\section{Statement of Intent}
This is a proposed standard that has neither been vetted nor ratified by any space agencies nor any government or public organizations. This is purely a demonstration of a more effective and pragmatic standard from the research perspective that can be worked on in the future to further develop it.

This proposed standard doesn't invalidate or question any of the previous work in this domain. The core purpose of this standard is to help further develop a more updated standard for the United Kingdom and ally state nations and space agencies to enhance security in satellite communication. Anyone from academia and the industry is highly welcome to take this as a reference or further work on this.

This proposed standard is a complete representation of the author's intellectual understanding based on analyzing the elements of satellite communications, threats, timeline of attacks, and various other criteria around satellite communication. Neither the Royal Holloway University of London nor the supervisor should be responsible for any forwards to this proposal.

The security architecture of Satellite Communication can have two kinds of recommendations – Recommendations for Standards and Recommendations for Practices. This proposed standard will provide guidelines on recommendations for practices due to the limitation of scope and time.

\section{Forward}
Questions relating to any of the areas or concerns of this document should be addressed with the author mentioned in the title of this document. The author is willing to cooperate and participate in further development of this proposed standard document.

\section{Document Control}
\begin{table}[h!]
\centering
\begin{tabular}{l l l l}
\toprule
\tabhead{Title} & \tabhead{Version} & \tabhead{Date} & \tabhead{Status} \\
\midrule
Standard for Secure Satellite\\ Communications & 1.0 & 20 August 2023 & Proposed\\
\bottomrule\\
\end{tabular}
\end{table}

\section{Purpose and Scope}
\subsection{Purpose}
This proposed standard document provides a very high-level guideline for developing the architecture of satellite communication taking the cyber security architectural viewpoints into serious consideration.

The goals of this proposed document are stated below.
\begin{itemize}
    \item Propose a high-level framework for integrating cybersecurity architecture in every segment of satellite communication.
    \item Provide guidelines in a high-level generic method so that it can be easily translated for diversified requirements in most satellite communication environments.
    \item Provide a base guideline for all stakeholders of satellite communication to achieve a higher level of security to better prepare and resist possible threats and attacks.
    \item Construct and progress the guideline in a modular fashion so it can be updated whenever a critical cybersecurity concern arises, and the framework required to accommodate it without completely changing massive segments.
    \item Develop guidelines based on generic risk management through the analysis of recent cyber-attack incidents on satellite communications.
    \item Establish zero-trust (\cite{noauthor_zero_nodate}) security architecture in every segment of satellite communication.
    \item Incorporate post-quantum-cryptography in the framework to keep the window open for incorporating futuristic cryptography.
\end{itemize}

\subsection{Scope}
The scope of this document will consider all possible kinds of satellite communication available today including the military, government, and private sectors. It will also include the satellite communication elements in the ground segment, user segment, and space segment. However, any other radio signal communication equipment or methods that directly or indirectly do not involve a satellite will be out of the scope of this proposed framework. This framework will provide a very high-level generic recommended practices for satellite communication. These may or may not be appropriate to be incorporated in other communication environments which are beyond the scope of this framework.

\section{Proposed Architecture}
This framework will follow the viewpoints stated in the CCSDS 351.0-M-1/BS ISO 20214:2015. However, this document will analyze each viewpoint and provide relevant additions and/or modifications to each of them.

\keyword{Enterprise Viewpoint} -- In addition to the mentioned elements involved in the organizational relationship, this document suggests considering suppliers of components in the supply chain of satellite communication architecture. Moreover, besides the organization, this document will also incorporate end users of satellite communications to provide better coverage for the stakeholders for secure satellite communication.

\keyword{Connectivity Viewpoint} -- In addition to the physical structure and physical environment of the space data systems, this document will suggest incorporating cyber elements for satellite communication. Moreover, it will also suggest considering portable and smaller ground stations (handheld receivers, transmitters, and modems) to ensure endpoint security in satellite communications.

\keyword{Functional Viewpoint} -- This document considers the functional viewpoint in the original document robust enough to be considered even in 2023.

\keyword{Information Viewpoint} -- This document considers the information viewpoint in the original document relevant enough to be considered even in 2023.

\keyword{Communication Viewpoint} -- In addition to the communication viewpoint mentioned in the original framework, the communication view should also describe the protocol stacks and mechanism of communication between any parties involved throughout the entire lifetime of the satellite.

\section{General Security Principles}
\subsection{General}
Security aspects of satellite communication architecture should include, but not be limited to, the aforementioned viewpoints described in section 5.2. This document suggests the following aspects.

\keyword{Physical Security} -- Enterprise Viewpoint, Connectivity Viewpoint.

\keyword{Information Security (Data in Rest)} -- Enterprise Viewpoint, Connectivity Viewpoint, Functional Viewpoint, Information Viewpoint, and Communication Viewpoint.

\keyword{Transmission Security (Data in Transit)} -- Connectivity Viewpoint, and Communication Viewpoint.

\subsection{Physical Security}
Physical security should refer to the physical elements in the overall satellite communication architecture. While developing the physical security architecture, it should consider the zero-trust architecture with explicit verification, least-privileged access, and assume breach. This document emphasizes zero-trust architecture on physical security because of the high availability of ground stations to a mass audience. With the availability of proof of concept of hacking one of the satellite ground stations (\cite{noauthor_starlink_nodate}), zero-trust architecture will significantly strengthen this specific guideline. Moreover, considering the recent advancement in Anti-Satellite (ASAT) weapons, ensuring the physical security of the space segment should be designed in such a way that increases the capability of detecting and avoiding physical attacks in any segment of satellite communication.

\subsection{Information Security}
The original document considered both data "at rest" and data "in transit" under the information security under information security. However, this document recommends considering data "at rest" and data "in use" under the Information Security section. Data "in transit" should be considered in the following "Transmission Security" section. As part of ensuring stronger security for data "at rest" and data "in use, the following methods can be considered.

\keyword{Data Classification} -- Data classification should be considered for the selection of techniques and controls to ensure higher satellite communication data security. Data can be classified on sensitivity, usage, value, and other criteria based on the requirements of a specific satellite communication environment.

\keyword{Data Federation} -- Due to a decrease in data quality while exported outside the environment (\cite{noauthor_extend_nodate}), this report recommends considering data federation. NHS UK Data Federation policy document (\cite{england_nhs_nodate}) can be used as a reference while developing the data federation mechanism for satellite communication. Satellite communication can involve data collection from multiple sources. Considering this, data federation can be considered as a tool for strengthening information security.

\keyword{Data Tokenization} -- Entities can consider replacing sensitive data with placeholder tokens. This requires less computing power than other mechanisms for data confidentiality which is useful considering the limited resources in different segments of satellite communication.

\subsection{Transmission Security}
Transmission security should be concerned with the data "in transit" and ensure confidentiality, integrity, and availability of data during the transmission of the data. This is crucial considering the recent (2000 - 2020) increase in attacks on satellite communication through advanced jamming techniques, capabilities, and reports of attacks.

\subsection{Procedures}
This document considers the procedures in the original document robust enough to be considered even in 2023.

\subsection{Mission Security Documentation}
\tocless\subsubsection{General}
Apart from the mentioned security policy and topic-specific policies, the following topic-specific policy should be considered.
\begin{itemize}
    \item Awareness, Training, and Disciplinary Procedures
    \item Disaster Recovery Policy
\end{itemize}

This carries significant weight for multiple reasons. First of all, according to the UK Government statistics, 83\% of the cyber-attacks were initiated through phishing attempts (\cite{noauthor_cyber_nodate-1}). Moreover, the Information Commissioner's Office emphasizes awareness and training (\cite{noauthor_training_nodate}) to become more resilient to cyber-attacks.

\tocless\subsubsection{Security Policy}
In addition to the given guidelines in the original document, this framework should regularly be reviewed and updated in accordance with the updates in the UK National Cyber Strategy (\cite{noauthor_national_nodate}) which is published at regular intervals. This framework security policy should be designed in such a way that it is possible to update it to follow, support, and comply with the UK national goals and cyber strategies in a timely manner. Moreover, the security policy should clearly document and specify achieving the goals and objectives through the zero-trust security architecture.

\tocless\subsubsection{Security Interconnection Policy}
In addition to the given guidelines in the original document:
\begin{itemize}
    \item Each participating stakeholder should be capable of connecting through protocols and controls under the zero-trust architecture implemented in the overall architecture of satellite communication.
\end{itemize}

\tocless\subsubsection{Risk Assessment}
In addition to the given guidelines in the original document, while doing the risk management, the BS ISO 27005:2022 should be considered for effective risk management in the satellite communication environment. Moreover, the discussion in Chapter 3 of this report should be taken into consideration to identify attack vectors and their possible level of impact.

\tocless\subsubsection{Mission Security Architecture}
This framework slightly modifies the guidelines in the original document. The mission security architecture should be designed with an emphasized focus on the security of the overall satellite communication.

\tocless\subsubsection{Security Operating Procedures}
The security operating procedures should strongly maintain the training and awareness policy of the framework besides the other relevant policies and should strongly maintain the zero-trust security architecture. The architecture should be modular and robust enough so it can accommodate the best possible procedures and technology based on the requirements without requiring additional budget or can come up with a solution that remains positive at the cost-benefit analysis.

\section{Security and the Reference Architecture}
\subsection{Overview}
In the current stage, this framework follows the CCSDS reference architecture as described in the original document. However, this leaves the window open for further research for coming up with a new reference architecture considering and keeping the highest interest of the United Kingdom and ally nation-states.

\subsection{Security and the Enterprise View}
\tocless\subsubsection{General}
In addition to the guidelines stated in the original document, this framework recommends that each and every data endpoint should follow the zero-trust security architecture, and this should be clearly documented in the main security policy of the organization. All participating stakeholders must adhere to the security policy of the main organization before getting any access to any data endpoint.

\begin{figure}
    \centering
        \includegraphics[scale=1]{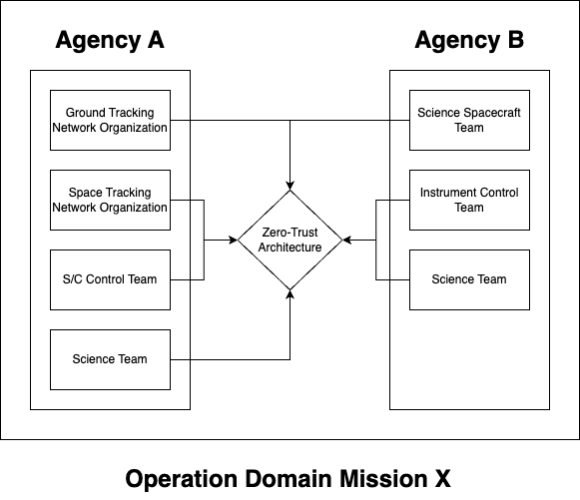}
        \caption{Proposed Enterprise View of Satellite Communication and Space Data Systems Security}
        \label{fig:enterprise_view}
\end{figure}

Moreover, this framework would recommend the following architectural view \ref{fig:enterprise_view} instead of the view [Figure 4-1] in the original document.

\tocless\subsubsection{Security Risks Highlighted by the Enterprise View}
This report proposes a completely different framework for the Enterprise View of satellite communication. The original document considered both cases where the participants trust each other and do not fully trust each other. The framework relied and guided on the intrinsic trust between the parties. However, this report will provide guidance on a zero-trust security architecture where in every layer, it will be assumed a breach, and the flow should proceed considering a breach has happened to tighten the security.

The original document suggested if the participants do not fully trust each other then they can either focus on the infrastructure or on the data. When focusing on the infrastructure, the main system should be isolated from all other untrusted systems to limit the damage. The original report also specified that this architecture is expensive and limits the mission data processing. When focusing on the data, the original document suggested putting a strong barrier between the system and the untrusted participants.

However, this report guides on the opposite. It strongly recommends implementing a zero-trust security architecture. In contrast with the classic security architecture, zero-trust recommends protecting all participating systems to be connected through a central policy mechanism assuming breach and implementing explicit verification with the least privileged access for all systems and components. This may seem expensive but compared to the opportunity cost of data breach, it justifies the cost-benefit analysis in the long run. Moreover, with the increased availability of satellite communication services, endpoints are getting more and more diversified and disbursed. In this widely disbursed environment, zero-trust security architecture will bring more benefit in terms of security compared to the classic security approaches.

\begin{figure}
    \centering
        \includegraphics[scale=0.9]{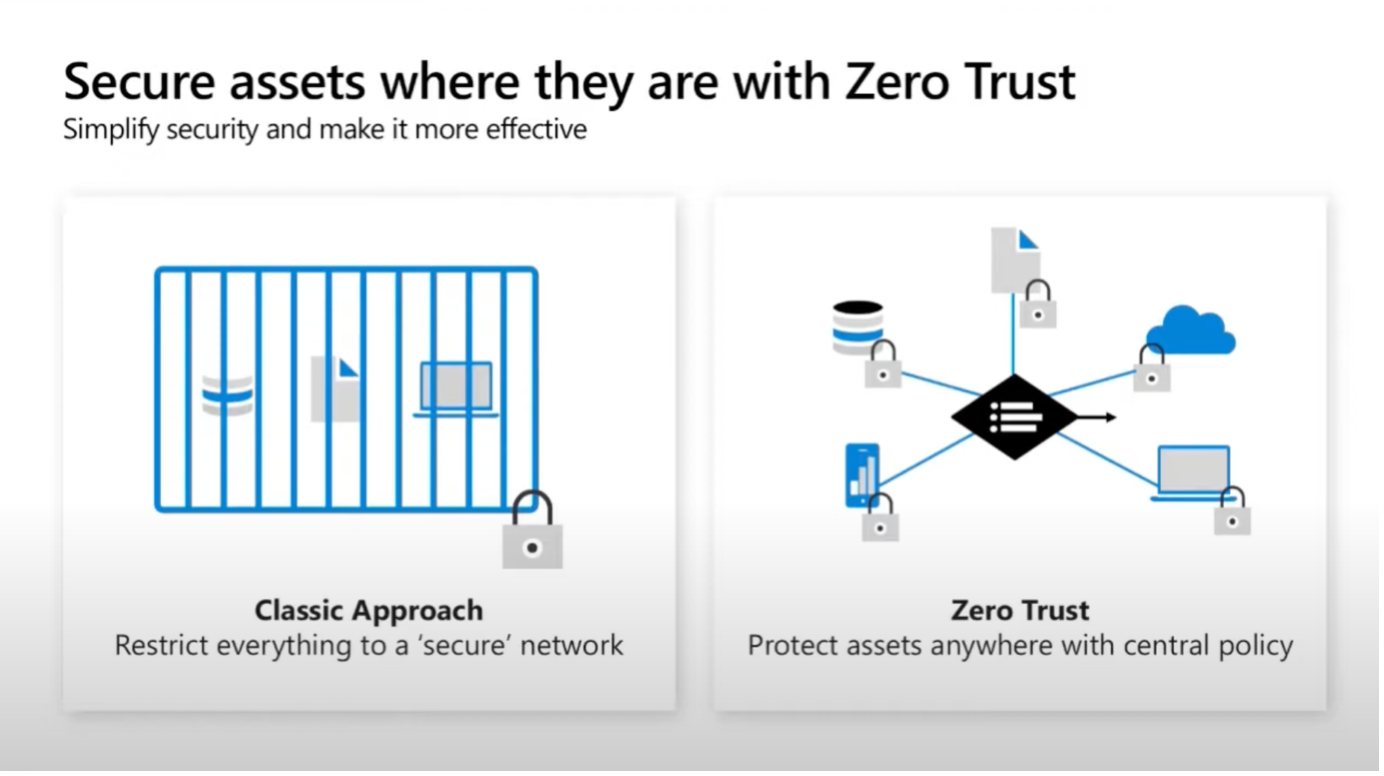}
        \caption{Zero-Trust compared to Classic Security Approach. Source: Microsoft (\cite{noauthor_mcra_nodate})}
        \label{fig:zero_trust}
\end{figure}

\subsection{Security and the Connectivity View}
\tocless\subsubsection{General}
The original document focused on ground-based connectivity and provided guidelines for that in a dedicated sub-topic [4.3.2]. In addition to that, this framework recommends also considering security for space-based connectivity. With the recent advancement of satellite communication, we now have more complicated satellite constellations and satellites communicate with each other within and beyond their own network. An attack can be initiated from another compromised satellite within the network or from a foreign satellite network in space. That is why space-based connectivity should also be considered.

\tocless\subsubsection{Ground Systems}
This framework agrees with the original document in the point that it guided the implementation of sufficiently robust controls to protect the system. However, the original document suggested using technologies like VPN and private operational circuits. This framework recommends using a zero-trust architecture with explicit verification. Theoretically, VPN networks provide higher security than no VPN. However, if the node, which is connecting to the system through VPN, gets compromised then the overall security of the system is at risk. With zero-trust architecture, even if a node gets compromised, the attacker has to explicitly verify the identity and the right to access any data or systems. This is more effective than trusting all the nodes and endpoints within the connectivity network or VPN by default. Zero-trust also helps in ensuring no default access to any component in the system.

\tocless\subsubsection{Space Systems}
With the recent advancement of satellite communication and satellite constellations, there are a series of satellites flying high above which are parts of a single network forming a mega-constellation (\cite{jonathan_amos_spacex_nodate}). These satellites are connected to each other at the space segment. Any compromised node can attack other satellites within or beyond its own network. Because of this architecture, risks in the space segment should be assessed and relevant security controls should be implemented to mitigate threats from within the space segment.

\tocless\subsubsection{Physical Security}
As mentioned in the original document, complete physical security is beyond the scope of this framework. However, the previous discussion on ASAT weapon development and the risks associated with it, this framework recommends taking pragmatic steps to ensure physical security both at the ground segment and the space segment in addition to the guidelines stated in the original document.

\tocless\subsubsection{Security Risks Highlighted by the Connectivity View}
In addition to the guidelines and risks suggested in the original document, this framework will recommend also considering unauthorized connectivity to the system through social engineering and other adversarial methods.

\subsection{Security and the Functional View}
\tocless\subsubsection{General}
This report completely agrees with the original document on the guideline that security should be considered from the outset of the overall mission design of satellite communication. However, the original document defines the changes as a "trade-off" while this framework considers finding better alternatives in the long run, in terms of security and functionality. Organizations should not compromise any security features or functionality for each other and consider this as a tradeoff. Rather they should research and develop a solution that fulfills all the requirements from both security and functionality perspectives.

\tocless\subsubsection{Security Risks Highlighted by the Functional View}
In addition to the guidelines mentioned in the original document, the functional view should also account for developing a proper policy for the zero-trust mechanism that doesn't hinder any required functionality within the system and the mission.

\subsection{Security and the Information View}
\tocless\subsubsection{General}
This document considers the information viewpoint in the original document relevant enough to be considered even in 2023.

\tocless\subsubsection{Risks Highlighted by the Information View}
\paragraph{General}
The original document covered all the basic security service requirements. This report finds that relevant enough to be considered even in 2023.

\subsection{Security and the Communications View}
This framework recommends considering every endpoint for assessing and describing the layered protocols and stacks involved in communication in contrast with the original document. The original document focused only on the network nodes. However, due to the operation in a broadcast nature, even the handheld receiver devices of the end users should also be considered for securing the overall satellite communication system.

Moreover, this framework recommends assessing the ability to implement post-quantum cryptography based on the requirements of security services in the physical layer. The system should be designed in such a way so that it can incorporate post-quantum cryptography mechanisms when available to be resilient against high-powered computational attacks on satellite communication.

This report finds the other guidelines relevant enough to be considered even in 2023.

\section{Security Architecture Principles}
\subsection{Overview}
This section will highlight the key security principles for satellite communication system designs.

\subsection{Open Standards}
The original document highly recommended prioritizing freely and easily available technologies. However, this framework recommends selecting appropriate technologies to secure required services through requirement engineering, regardless of whether they are freely available or not. This framework suggests selecting technologies that follow all 6 of Kerckhoffs's principles (\cite{kerckhoffs_cryptographie_1883}) so that even if everything except the private key or key generating keys or any other sensitive secret keys about the technology gets publicly disclosed, that should not have any impact on the security of the overall systems. If this requires using proprietary technologies, the organization should utilize them based on the sensitivity of the data. However, the pros and cons of using proprietary technologies should be assessed as discussed by Professor Keith Martin in his book Everyday Cryptography (\cite{martin_everyday_2017}). This framework agrees with other recommendations in the original document, for example, the technology should be selected in such a way that it is compatible with the participants.

\subsection{Protection throughout Layered Security Mechanisms}
This framework suggests the same multiple layers of security which is one of the core concepts of zero-trust architecture.

\subsection{Expandability}
This document considers the guidelines in the original document relevant enough to be considered even in 2023.

\subsection{Flexibility}
This document considers the guidelines in the original document relevant enough to be considered even in 2023.

\subsection{Interoperability}
Besides the guidelines provided in the original document, this framework suggests interoperating with other organizations through the zero-trust policy as the baseline standard. This document considers the other guidelines in the original document relevant enough to be considered even in 2023.

\subsection{Key Management}
This document considers the guidelines in the original document relevant enough to be considered even in 2023.

\subsection{Encryption Algorithm Selection}
In contrast with the original document, this framework recommends emphasizing the security of the mission over interoperability. However, this framework does not indicate that interoperability can be avoided for the sake of security. The organization should assess and select the best cryptographic mechanism to ensure higher security that still provides interoperability for the success of the overall mission.

\subsection{Kerckhoff's Principle}
In contrast with the original document, not only the cryptosystems but also all other technologies and controls within the overall system should try to maintain Kerckhoff's principles as suggested in section 5.5.2.

\subsection{Fault Tolerance}
Besides the guidelines in the original document, this framework also suggests creating a Disaster Recovery sub-topic policy as mentioned in section 5.3.6.1. The organization should consider a fallback and secret communication channel and/or technology at the space segment so that in case of jamming and other ground-based attacks, it can still be operated from other satellites to recover from the disaster. In contradiction with the original document, this framework recommends finding recovery methods without degrading the security mechanisms as that can introduce new threats in the future.

\section{Mission Profiles}
\subsection{Overview}
In addition to the mission profiles discussed in the original document, this framework would like to include the following under the "Scientific" mission profile due to the advancement in this segment in recent years and due to the potential to advance more in this segment in the upcoming years.

\begin{itemize}
    \item Space Observation
\end{itemize}

Moreover, this framework suggests focusing on duplex communication through satellites considering the recent advancements of satellite usage by generic audiences in duplex mode (e.g., Starlink Internet through Satellite Communication) under the "Communications" mission profile.

\subsection{General}
In addition to the provided guideline in the original document, this framework suggests conducting an assessment of possible fallback approaches in case of an incident without compromising security level.

\subsection{Human Spaceflight}
Because of the sensitivity of the operation mentioned in the original document, additional communication channels should be considered in case of an attack or unavailability of the primary communication channel for human spaceflights.

\subsection{Earth Observation}
This framework agrees with the mentioned importance of the Earth observatory satellites. This framework suggests considering that to a significant extent because any anomalies in earth observation data may or may not have an immediate impact, but observation data of longer periods are used for prediction and many other purposes. So, ensuring the integrity and availability of the data holds critical importance.

\subsection{Communications}
This framework will suggest adding quite a few guidelines in addition to the mentioned ones in the original document. First of all, the original document mentioned that communication systems are usually based on geostationary satellites. Though it may have been the case in 2012, satellite communications now heavily involve LEO satellites. Hundreds of LEO satellites are now being used to provide duplex high-speed high-bandwidth communication to provide data connectivity even in remote areas. The deployment of a constellation of communication satellites in 2012 is now covering the whole earth in 2023.

\begin{figure}
    \centering
        \includegraphics[scale=1]{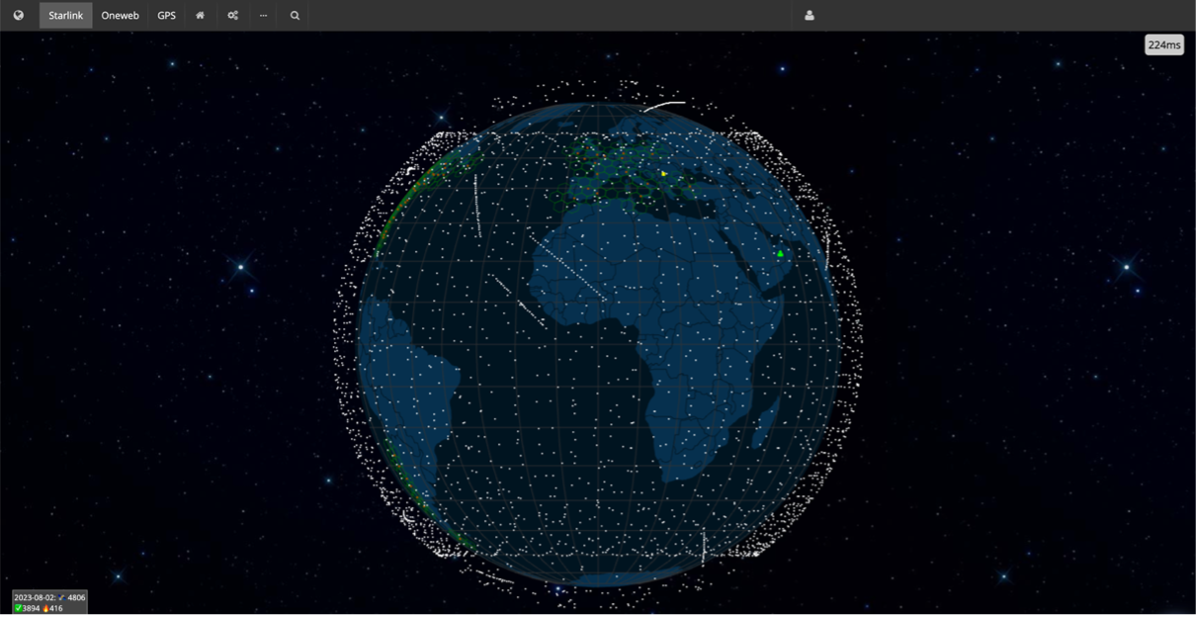}
        \caption{Starlink Satellite Constellation Live Map. Source: SatelliteMap.space (\cite{noauthor_starlink_nodate-1})}
        \label{fig:starlink}
\end{figure}

The original document emphasized GEO satellites because of their extensive coverage on Earth. However, as shown in Figure 17, LEO satellite constellations have near full coverage of Earth. The original document considered potentially reduced threats to LEO satellites because of their shorter visibility but those satellites are connected to the other LEO satellites in the constellations to a greater extent which is also acknowledged in the original document. That is why this framework suggests considering both GEO and LEO with equal importance and assessing the risks around all kinds of satellite constellations based on the requirements of the mission thus increasing resilience against probable threats.

\subsection{Scientific}
\tocless\subsubsection{Near Earth Orbit}
In addition to the original guidelines, this framework suggests also considering the satellite-to-satellite communication security at the space segment besides the previously mentioned telecommand channel in the original document.

\tocless\subsubsection{Lunar}
This document considers the guidelines in the original document relevant enough to be considered even in 2023.

\tocless\subsubsection{Interplanetary/Deep Space}
This document considers the guidelines in the original document relevant enough to be considered even in 2023. As an example, on August 1, 2023, NASA lost contact with the Voyager 2 after accidentally sending a wrong command that moved the antenna of Voyager 2 two degrees off. This resulted in lost communication (\cite{marianne_guenot_nasa_nodate}). However, Voyager 2 has a pre-program that will reset the antenna direction after a certain period of time. This proves that the guidelines in the original document are still relevant, even in an accidental case in 2023.

\subsection{Navigation}
Besides the guidelines in the original document, this framework will recommend deploying zero-trust architecture while delivering data to vehicles depending on satellite communication. On the Royal Holloway MSc. Information Security program, under the Cyber Crime module, a demonstration was given regarding how much loss a compromised navigational command can cause. If the receiver device doesn't verify the data received from the satellite, and if the data contains malicious commands, it can block critically important channels can cause millions of pounds of damage every single day. Incorrect navigational data in aircraft and air-based weapons can cost the lives of civilians. That is why it is utterly crucial to implement the zero-trust security architecture at the ground segment where the end-users are in the satellite communication environment. The satellite agencies should ensure that the receiving technology complies with the policies and guidelines of the zero-trust architecture just like any other components in the satellite communication system.

\subsection{Multi-organizational Spacecraft}
This framework will skip this sub-section as this is a specific sub-profile under the aforementioned profiles.

\section{Proposed Architecture}
\subsection{Requirements}
In addition to the requirements stated in the original document, this framework would like to propose the following requirements:

\begin{itemize}
    \item The architecture should consider integrating zero-trust security architecture in all layers of functions within the system.
\end{itemize}

The other requirement in the original document actually supports the guidelines provided in this proposed framework.

\subsection{Services}
In addition to the mentioned services in the original document, this framework suggests the following services.

\textbf{Security Services:}
\begin{itemize}
    \item Data availability during the entire mission.
\end{itemize}

\textbf{Security Mechanism:}
\begin{itemize}
    \item Periodic Stakeholder Training and Awareness (to ensure resilience against social engineering and other human-centric attacks).
\end{itemize}

\textbf{Consideration on the implementation of security mechanism:}
\begin{itemize}
    \item Explicit Verification.
    \item Least-privileged Access
    \item Assumed Breach
\end{itemize}

\subsection{Proposed Security Architecture}
This framework proposes a modified version of the CCSDS Security Core Suite which is amended in the following sections.

\subsection{Satellite Communications Security Core Suite}
\tocless\subsubsection{Aims of the Security Core Suite}
The modifications and the differences between this framework and the CCSDS Security Core Suit are stated below.
\begin{itemize}
    \item Security mechanisms should follow the zero-trust architecture, so it doesn't require combining the sensitive stack in a central environment and putting emphasized security on that central environment. Rather, every element within the whole system should follow the zero-trust policy, go through explicit verification for any communication, and have least-privileged access to the system.
    \item The application of protocols is beyond the scope of this framework and keeps the window open for further research into this area.
\end{itemize}

\tocless\subsubsection{Security Core Suite Definition}
This framework suggests that security services should be applied in each layer and in each endpoint in each component through the zero-trust central policy. It doesn't exclusively recommend any specific mechanism. It encourages utilizing the zero-trust principles while developing the core policy, so the final system maintains the requirements of the zero-trust security architecture. The choice of services should be based on the requirements of the mission.

It follows the operational combinations mentioned in the table under section 7.4.2 in the original document.

\subsection{Security Core Suite Configuration}
\tocless\subsubsection{General}
This document considers the guidelines in the original document relevant enough to be considered even in 2023.

\tocless\subsubsection{Data Link Layer}
This document considers the guidelines in the original document relevant enough to be considered even in 2023.

\tocless\subsubsection{Network Layer Security}
This document considers the guidelines in the original document relevant enough to be considered even in 2023.

\tocless\subsubsection{Application Layer Encryption}
This document considers the guidelines in the original document relevant enough to be considered even in 2023. However, the original document guided integrating implementation libraries, but this framework will add to that by suggesting conducting a robust security assessment of the library before integrating it into the system and doing the same before updating the version of each of the libraries to ensure supply chain security.

\tocless\subsubsection{Payload Specific Security}
In contrast with the original document, this framework suggests only using the security mechanisms that support the zero-trust architecture and can be integrated through a zero-trust policy.

\subsection{Expandability}
In addition to the guidelines in the original document, this framework suggests only selecting security mechanisms that comply with the zero-trust architecture in every segment of the overall system.

\subsection{Emergency Operations}
This framework would like to add to the guidelines in the original documents from a different perspective.

The original document provided guidelines for considering three levels around security – full authentication, reduced authentication, and non-authenticated commands. That also provided a sequence for this. However, this framework strongly recommends not accepting any kind of non-authenticated commands. Rather, during the mission design, the architecture should be designed in such a way that incorporates sending mission-critical control commands from other sources to the satellite under full authentication.

In the original document, an example scenario is provided where a spacecraft may have a failed command and control system and that may lead to a point where non-authenticated commands need to be sent to the spacecraft. However, this framework considers this as a major security risk. Adversaries can forcefully put the spacecraft into emergency mode by deliberately attacking the communication system of the spacecraft and then utilize this weakness to take further control of the spacecraft and use it for adversarial purposes. A compromised spacecraft can not only become a threat to its own communication systems but also become a threat to other spacecraft and systems.

Because of this possibility, this framework strongly recommends not accepting any kind of commands without explicit verification and proper authentication.

The original document also mentioned that in some cases, it may require sending control signals in clear mode but that leaves the attack window open for advanced adversaries. Again, this framework will strongly recommend not accepting any kind of control commands without proper authentication and authorization. The system should be designed in such a robust way that it should have an alternative emergency communication protocol with proper authentication and authorization. In case of anomalies, the system should have the capability to correct itself based on the pattern of previous commands. This is possible as seen on the Voyager 2 space probe that was launched in 1977. In the recent event of lost communication, Voyager 2 will correct its antenna position without any additional control commands after a certain period of time. This proves that it is possible to design systems in the way this framework recommends.

\section{Challenges}
This report did a PESTLE analysis on the challenges and barriers to acceptance and implementation of the proposed standard. The analysis report is provided below.

\textbf{Political} -- All organizations operate under and directly or indirectly rely on the political environment. The impact of changes in political views and environments was considered while developing the proposed standard.

The proposed standard does not reflect or oppose any specific political view and vision. Rather, it provides a very high-level guideline for achieving higher security for satellite communications and space data systems. Political environments may obstruct or delay the cooperation from different state nations in accepting and implementing the proposed standard. However, the proposed standard itself doesn't have any impact on the political environment. It supports the current objective of the United Kingdom government to achieve higher excellence in satellite communication. Because of the alignment with current government objectives, the political environment should not appear as a barrier to accepting and implementing the proposed standard.

\textbf{Economic} -- The success of any mission heavily relies on the economic feasibility.

The proposed standard does not propose a completely different approach than what is being practiced in the industry at this moment. Rather, it suggests strategies to implement better security approaches. Implementing the new approach will require an additional budget. Because of the nature of the satellite communication architecture, the budget can become significantly large. However, comparing the required budget with the security benefits it will bring and the loss that occurred through attacks makes this proposal pragmatic and required. The updates do not require recurring budgets. Once the newer strategies are implemented, it can provide the benefits until some ground-breaking threat landscape appears and completely new strategies are required for that.

\textbf{Social} -- Social and cultural environments can have a reasonable impact on the success of an operation if the operation is heavily connected with the aforementioned environment.

The proposed standard does not have much impact on the social and cultural environment. It does not propose any major changes in how people live in society and run through the culture. That is why the proposed standard will not have considerable challenges from the social and cultural environment.

\textbf{Technology} -- Technologies are one of the crucial concerns for the majority of the operations in the world today.

The proposed standard provides updates in strategies that will require the implementation of new security technologies in different segments and elements of the satellite communications and space data systems. This is a reasonable challenge for the proposed standard. However, stakeholders can slowly upgrade their technological controls as this does not obsolete the current technological architecture and elements. Rather, it suggests accommodating newer technologies with the existing architecture. That is why this is a significant challenge but not impossible to achieve in the long run.

\textbf{Legal} -- Every operation should run abiding by the legal and regulatory requirements.

The proposed standard does not violate any legal and regulatory requirements. Rather, it strengthens the confirmation of following legal requirements by offering stronger information security. That is why there should not be any legal or regulatory challenges for acquiring and implementing the proposed standard.

\textbf{Environmental} -- The environmental impact of any operation should be considered with a high priority.

The proposed standard does not propose anything that may become a threat to the environment. Rather it proposes strategies for securing satellites that may result in reducing space debris. That is why there are no environmental challenges to the proposed standard.

\subsection{PESTLE Summary}

From the theoretical analysis, we can see that the proposed solution will face a low level of political, medium level of economic, low level of social, high level of technological, and low level of environmental challenges. The synopsis of the PESTLE analysis is presented in table \ref{tab:pestle_analysis}. The overall challenges are on the lower medium range which makes it a pragmatic strategic solution for securing satellite communications and space data systems.

\begin{table}[h!]
\centering
\begin{tabular}{l l l l l l}
\toprule
\tabhead{Political} & \tabhead{Economical} & \tabhead{Social} & \tabhead{Technological} & \tabhead{Legal} & \tabhead{Environmental} \\
\midrule
Low & Medium & Low & High & Low & Low\\
\bottomrule\\
\end{tabular}
\caption{Summary of PESTLE Analysis}
\label{tab:pestle_analysis}
\end{table}

\section{Scope for further research}
The proposed standard provides very high-level guidelines for satellite communication and space data security. This introduces the opportunity to further research for developing topic-specific policies for space missions. It also keeps the windows open for research on specific technological approaches (e.g. Cryptography, Physical Layer Security, etc.) and systematization of the proposed standard.

\chapter{Conclusion} 

\label{Chapter7} 


\setcounter{secnumdepth}{5}

Satellites are an inseparable part of remote wireless communications today due to the significant coverage of the areas where traditional communication methods struggle to reach. From television program broadcasting to military communications, from remote wireless internet to emergency communication services, from maritime industry to earth and space observation, satellites play a significantly important role. Deployment of a satellite generally requires a huge budget but the availability of renting satellite capacity made it available even to smaller organizations and individuals. The satellite industry is relatively huge in financial terms and consolidated revenue is reported to be USD 159 Billion. This market is expanding with more rockets flying into space carrying multiple satellites in a single run. The governments of different state nations are also becoming more interested in security satellite communications. For example, the National Cyber Security Center of the United Kingdom organized a workshop on securing satellite communication which the author was privileged to be invited to.

Satellite communications can be divided into three major segments which are the ground segment, space segment, and the ecosystem. The space segment consists of the satellites and their constellations up in space. The ground segment consists of all the networks and systems on earth controlling and managing the communications. The ecosystem segments consist of all the technologies, mechanisms, engineering, architecture, and other associated concerns revolving around satellite communications. Based on the positional state and distance from Earth, satellites can be divided into GEO, MEO, LEO, SSO, and Lagrange Point satellites. This impacts their coverage on earth and availability from a specific point on Earth, also resulting in diversification in the threat landscape.

The journey started in 1957 with the launch of the first satellite with a radio transmitter. By the next few decades, satellites started being used for television broadcasting, amateur radio communications, commercial communications, remote weather sensing, satellite-to-satellite relay, global positioning, telephone service, and gigabit remote wireless internet connectivity in recent years. The technologies are being updated continuously and satellites are opening new windows of opportunities in different areas.

With the increase in the number of satellites and use of satellite communications, threat vectors, and attack landscape is also widening. The threats and attacks can be categorized into two broad categories - physical and cyber. Physical attacks can be further divided into two major sub-categories - kinetic energy attacks and direct energy attacks. Though theoretically possible, there has been no public report of these attacks being practiced. On the other hand, cyber attacks can be in terms of cryptographic, social engineering, eavesdropping, software threats, hardware threats, spoofing, unauthorized transponder usage and information modification, denial of service, and many other forms. This report has found proof of concept and reports of cyber attacks on satellite communications in the last few decades. The attack timeline shows a slow increase in cyber attacks on satellite communications over the last few years.

The study of satellite elements and segments, the historical growth of the satellite industry, its usage, also significant growth in the attack vectors for satellite communications and space data systems indicate the importance of combined forces from the military, government, and space agencies. As the requirements for each space mission has considerable differences and based on the position of the satellite and many other variables, only a technical solution is not enough to serve all the security purpose. The ideal approach is achieving resilience through a robust and updated security framework for satellite communications and space data systems. The effective approach to increasing resilience is through effective and pragmatic standards. Different state nations and space agencies have their own public and private standards. However, there is a significant gap in this domain for work. The relevant and publicly available standards were analyzed and found that many of them are for specific segments or elements of the satellite communication. The most comprehensive standard document is the BS ISO 20214 which was released in 2015 (8 years ago) and incorporated a CCSDS standard document that was released in 2012 (11 years ago). While most of the guidelines are still relevant, there is considerable scope for updating the guidelines.

This report analyzed the enterprise security architecture frameworks and security models and accommodated the ideology of the appropriate and relevant security models by proposing an updated standard for secure satellite communications and space data systems. Due to the integration with various endpoints in widely diversified environments, accommodating zero trust ideology can be highly effective and pragmatic. This report suggested the accommodation within the proposed updated standard. This will help both academia and the industry to use this proposed standard as a baseline for further research in this domain to achieve more robust and secure satellite communications and space data systems.


\appendix 


\chapter{Survey Data} 

\label{AppendixA} 

\section{Questionnaire}
\textbf{Question: Your Age?} [Select from the following options]

Options:
\begin{itemize}
    \item 15 - 25
    \item 25 - 35
    \item 35 - 45
    \item 45+
\end{itemize}

\textbf{Question: Your Highest Educational Qualification?} [Select from the following options]

Options:
\begin{itemize}
    \item Undergraduate
    \item Postgraduate
    \item PhD
\end{itemize}

\textbf{Question: Your Current Organization Name?} [Write inside an input field]

Options: Input Field

\textbf{Question: Your Field of Work/Study (e.g. Cybersecurity, Finance etc.)} [Write inside an input field]

Option: Input Field

\textbf{Question: Your Gender?} [Select from the following options]

Options:
\begin{itemize}
    \item Woman
    \item Man
    \item Non-binary
    \item Transgender
    \item Prefer not to say
\end{itemize}

\textbf{Which Country do you currently live in?} [Write inside an input field]

Option: Input Field

\textbf{Question: Have you traveled by air in the last 3 years?} [Select from the following options]

Options:
\begin{itemize}
    \item Yes
    \item No
\end{itemize}

\textbf{Question: Have you ordered any product that was shipped from overseas in the last 3 years?} [Select from the following options]

Options:
\begin{itemize}
    \item Yes
    \item No
\end{itemize}

\textbf{Question: Do you or any of your close neighbors have a satellite dish connected to their house for satellite TV?} [Select from the following options]

Options:
\begin{itemize}
    \item Yes
    \item No
\end{itemize}

\textbf{Question: In your understanding, how dependent you are on satellite communication? (1 - Not dependent at all, 10 - Severely dependent)} [Select from the following options]

Options:
\begin{itemize}
    \item 1
    \item 2
    \item 3
    \item 4
    \item 5
    \item 6
    \item 7
    \item 8
    \item 9
    \item 10
\end{itemize}

\textbf{Question: How badly do you think you will be impacted in case of a major attack on satellite communications? (1 - Not impacted at all, 10 - Severely impacted)} [Select from the following options]

Options:
\begin{itemize}
    \item 1
    \item 2
    \item 3
    \item 4
    \item 5
    \item 6
    \item 7
    \item 8
    \item 9
    \item 10
\end{itemize}

\textbf{Question: *ONLY* technological measures are enough to secure satellite communications} [Select from the following options]

Options:
\begin{itemize}
    \item Strongly Agree
    \item Agree
    \item Neutral
    \item Disagree
    \item Strongly Disagree
\end{itemize}

\textbf{Question: Governance and Compliance are required to secure satellite communication
} [Select from the following options]

Options:
\begin{itemize}
    \item Strongly Agree
    \item Agree
    \item Neutral
    \item Disagree
    \item Strongly Disagree
\end{itemize}

\textbf{Question: How important do you think "ethics" is for satellite communications?} [Select from the following options]

Options:
\begin{itemize}
    \item Extremely important
    \item Somewhat important
    \item Neutral
    \item Somewhat not important
    \item Not important at all
\end{itemize}

\textbf{Question: Militaries, Government, and Private Sectors of Different Countries should work together to develop standards in satellite communications} [Select from the following options]

Options:
\begin{itemize}
    \item Strongly Agree
    \item Agree
    \item Neutral
    \item Disagree
    \item Strongly Disagree
\end{itemize}

\textbf{Question: Governments should put more focus on securing satellite communications} [Select from the following options]

Options:
\begin{itemize}
    \item Strongly Agree
    \item Agree
    \item Neutral
    \item Disagree
    \item Strongly Disagree
\end{itemize}

\textbf{You are giving your consent for recording your responses in this form for the purpose of academic research under the Royal Holloway University of London.} [Select the option below]

Option:
\begin{itemize}
    \item Yes
\end{itemize}

\section{Responses}

\begin{figure}
    \centering
        \includegraphics[scale=0.6]{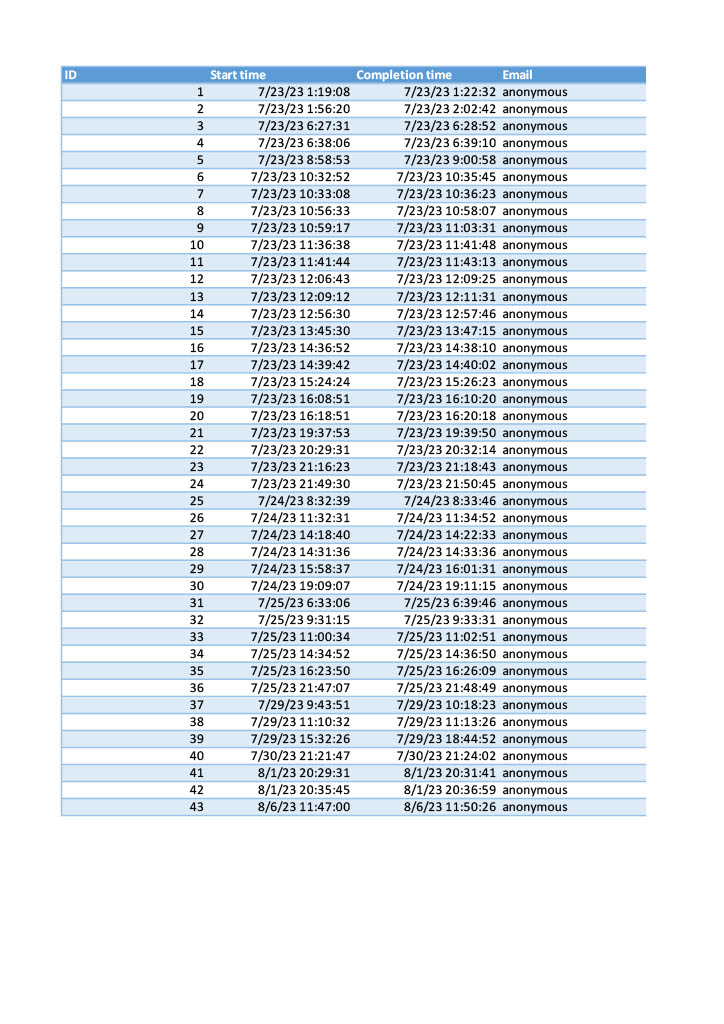}
        \caption{Survey Responses First Page}
        \label{fig:survey_1}
\end{figure}
\begin{figure}
    \centering
        \includegraphics[scale=0.6]{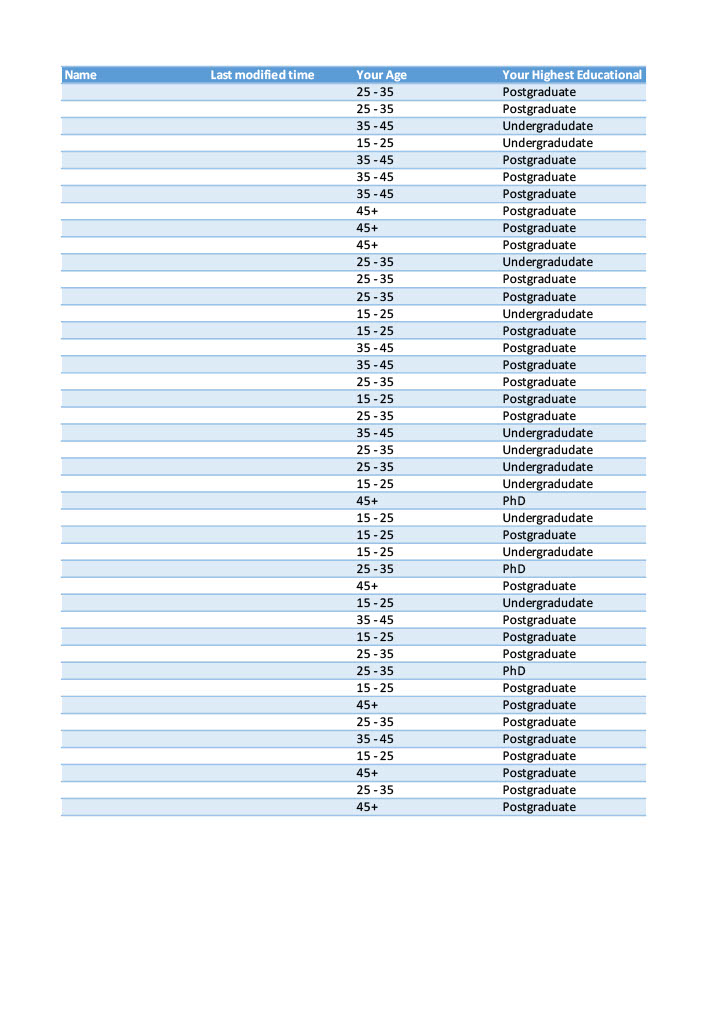}
        \caption{Survey Responses Second Page}
        \label{fig:survey_2}
\end{figure}
\begin{figure}
    \centering
        \includegraphics[scale=0.6]{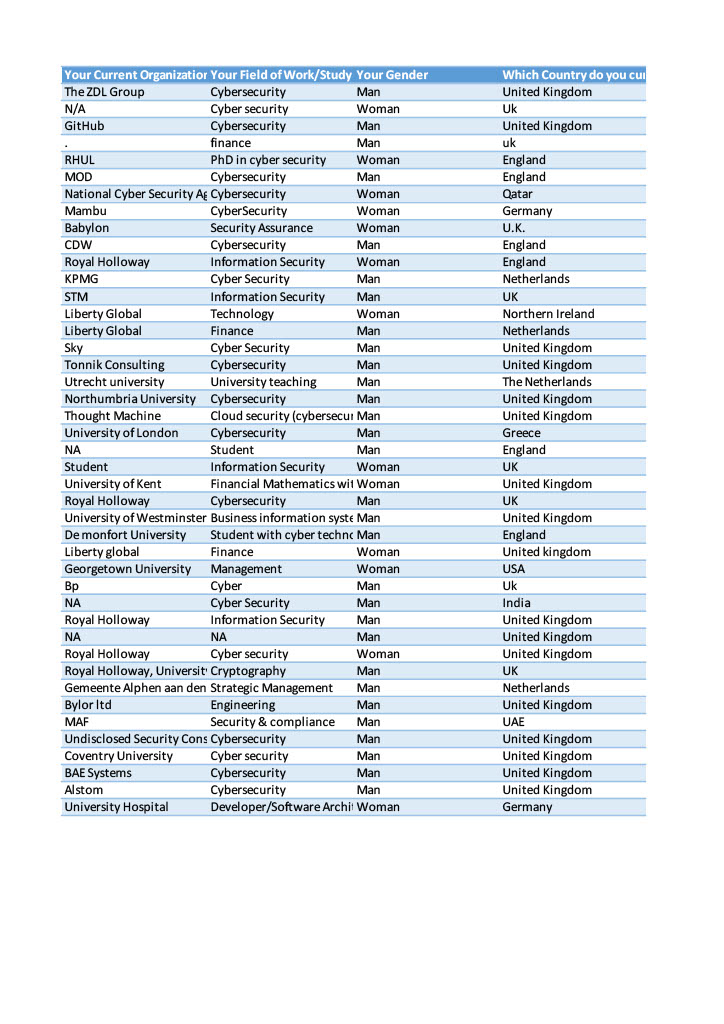}
        \caption{Survey Responses Third Page}
        \label{fig:survey_3}
\end{figure}
\begin{figure}
    \centering
        \includegraphics[scale=0.6]{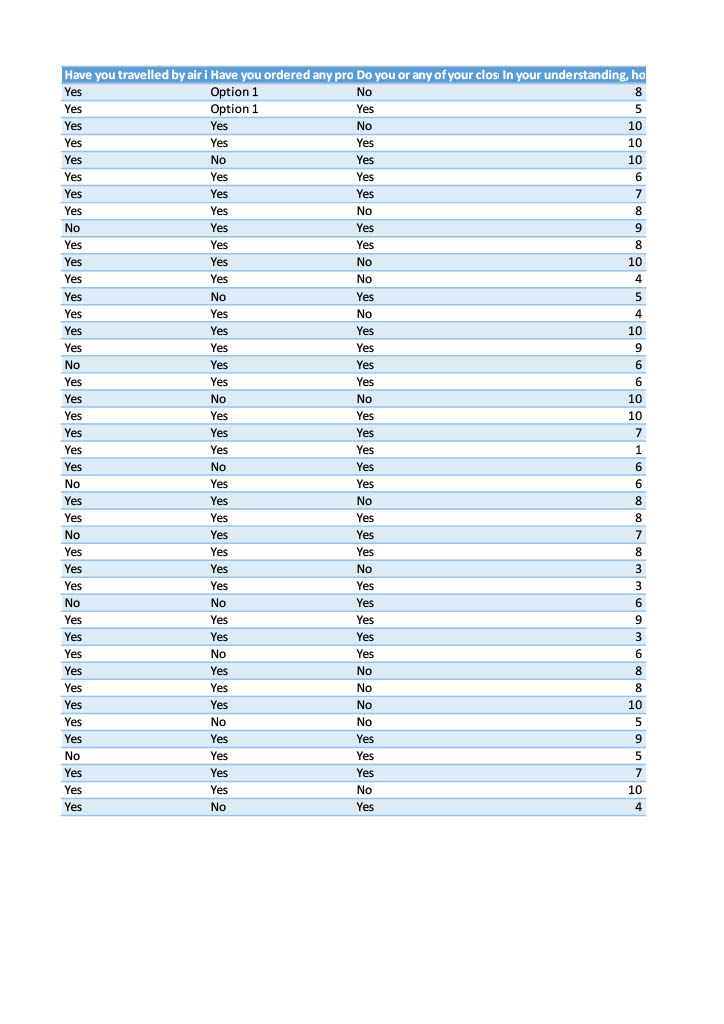}
        \caption{Survey Responses Fourth Page}
        \label{fig:survey_4}
\end{figure}
\begin{figure}
    \centering
        \includegraphics[scale=0.6]{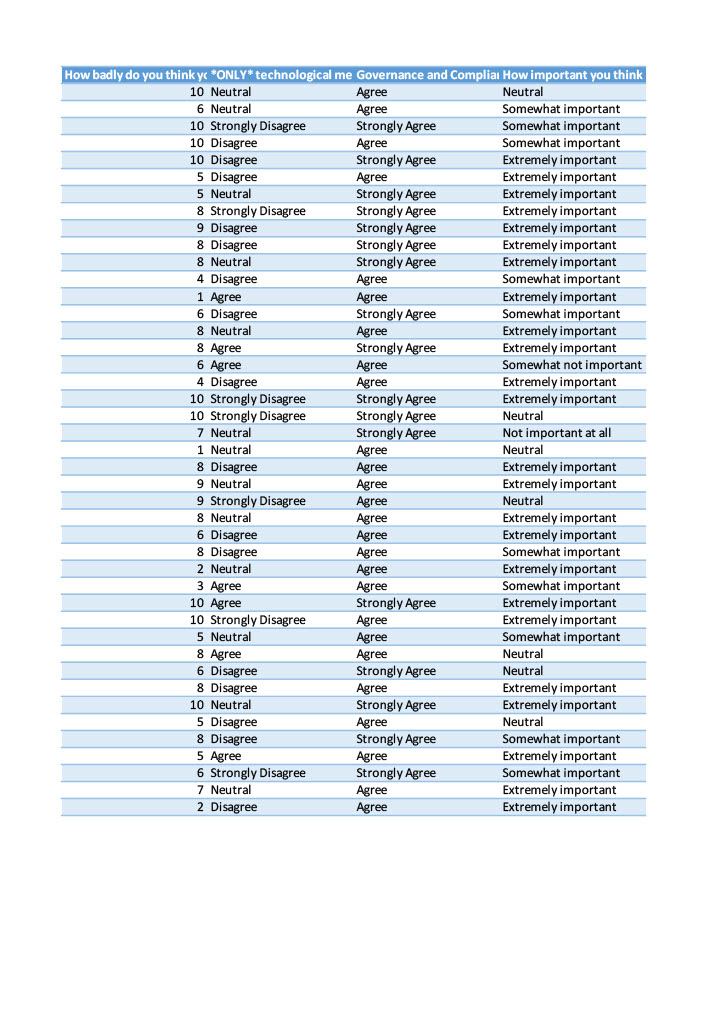}
        \caption{Survey Responses Fifth Page}
        \label{fig:survey_5}
\end{figure}
\begin{figure}
    \centering
        \includegraphics[scale=0.7]{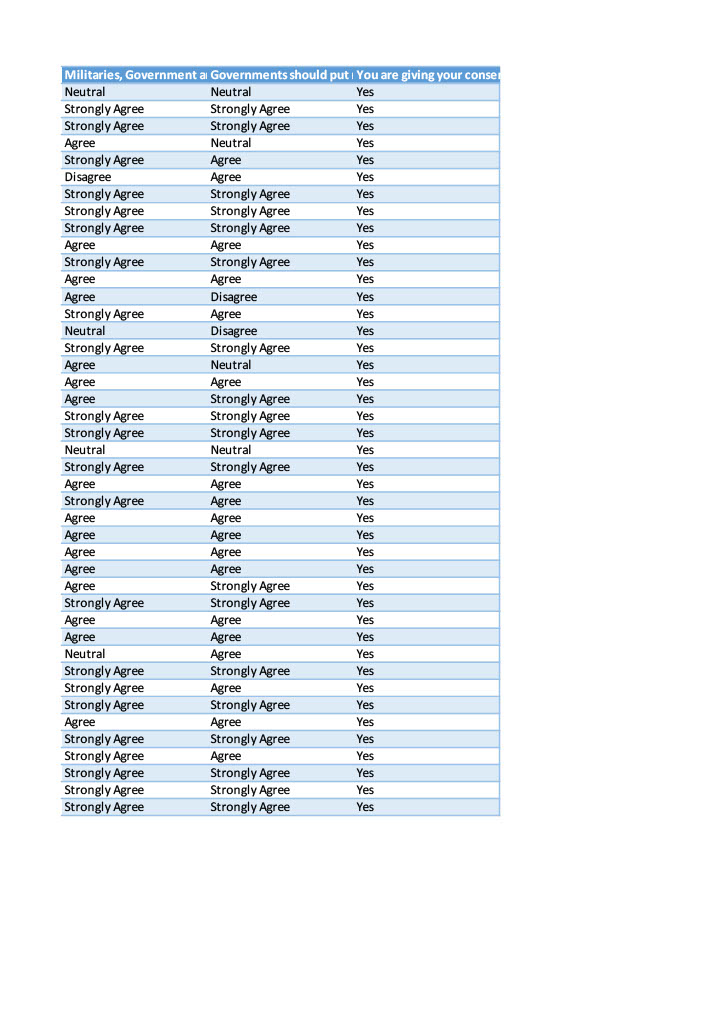}
        \caption{Survey Responses Sixth Page}
        \label{fig:survey_6}
\end{figure}


\printbibliography[heading=bibintoc]


\end{document}